\documentclass{article}
\usepackage[T1]{fontenc}
\usepackage[latin1]{inputenc}
\usepackage{graphics}
\usepackage{graphicx}
\usepackage{longtable}
\usepackage{color}
\usepackage{pslatex}
\usepackage{hyperref}
\usepackage{sverb}
\usepackage{subfigure}

\usepackage{a4p}
\usepackage{rotating}

\definecolor{airforceblue}{rgb}{0.36, 0.54, 0.66} 
\definecolor{amaranth}{rgb}{0.9, 0.17, 0.31}
\def\MB{\color{black}}
\def\ZW{\color{black}}
\begin{document}
\begin{titlepage}
 
\begin{flushright} 
{ IFJPAN-IV-2018-3 
} 
\end{flushright}
 
\vskip 3 mm
\begin{center}
{\bf\huge  Documentation of {\tt TauSpinner} algorithms -- program for simulating spin effects
in $\tau$-lepton production at LHC}
\end{center}
\vskip 13 mm

\begin{center}
   {\bf T. Przedzinski$^{a}$, E. Richter-W\c{a}s$^{a}$ and Z. W\c{a}s$^{b}$  }\\
       {\em $^a$ Institute of Physics, Jagellonian University,  30-348 Cracow, Lojasiewicza 11, Poland} \\
       {\em $^b$  Institute of Nuclear Physics, Polish Academy of Sciences,
        31-342 Cracow, ul. Radzikowskiego 152, Poland}\\ 
\end{center}
\vspace{1.1 cm}
\begin{center}
{\bf   ABSTRACT  }
\end{center}

The $\tau$-lepton plays an important role in the physics program at the Large Hadron Collider 
(LHC). It offers a powerful probe in searches for New Physics and can be used to measure parameters of the Standard Model.
Spin of $\tau$ lepton represents an interesting phenomenological quantity which can be used
for the sake of separation of signal from background or in measuring  properties
of particles decaying to $\tau$ leptons. 
A proper treatment of $\tau$ spin effects in the Monte Carlo simulations is important for
understanding  the detector acceptance as well as for the measurements/use of $\tau$
polarization and $\tau$ spin correlations in experimental analysis.

The {\tt TauSpinner} package represents a tool which can be used to modify $\tau$ spin effects in any
sample containing $\tau$ leptons. Also production matrix elements can be modified with its help.
The general principle of {\tt TauSpinner} algorithm is to rely on the kinematic of
outgoing particles (or jets) only, and to average over all possible initial
states accordingly to assumed parton distributions and cross-sections.
Incoming on mass-shell parton four-momenta are reconstructed in approximation,
but energy-momentum conservation is obeyed.
Generated samples of events featuring $\tau$ leptons with their decay products,
 can be used as an input.
The information on the polarization and spin correlations, and production/decay matrix elements
 is reconstructed from the kinematics of the 
$\tau$ lepton(s) (also $\nu_\tau$ in case of $W$-mediated processes) and $\tau$ decay products (also jets). 
No other information stored in the event record is required, except kinematics of these final state particles.
By calculating spin (or production/decay matrix element) weights, attributed on the event-by-event basis, 
it enables numerical evaluation of modifications due to these effects without the need for regenerating 
events.
With {\tt TauSpinner}, the experimental techniques developed over years since LEP 1 times are already 
used and extended for LHC applications.

Several applications of {\tt TauSpinner} were prepared. Elements
of program functionalities  were presented in appendices of consecutive publications together with 
explanation how the program can be used in each case.  The purpose of present publication is to 
systematically document physics basis of the program, as well as to give short overwiew of its application domain.

\vskip 1 cm


\vspace{0.2 cm}
 
\vspace{0.1 cm}
\vfill
{\small
\begin{flushleft}
{   IFJPAN-IV-2018-3
\\ February 2018
}
\end{flushleft}
}
 
\vspace*{1mm}
\footnoterule
\noindent
{\footnotesize \noindent  $^{\dag}$

This project was supported in part from funds of Polish National Science
Centre under decisions UMO-2014/15/B/ST2/00049 and  by the Research Executive  Agency (REA) 
of the European Union under the Grant Agreement PITNGA2012316704 (HiggsTools).

Majority of the numerical calculations were performed at the PLGrid Infrastructure of 
the Academic Computer Centre CYFRONET AGH in Krakow, Poland.

}
\end{titlepage}

\section{Introduction}

Physics of $\tau$ leptons at the LHC is oriented towards use of $\tau$ lepton decays 
for the measurement of properties  of hard production processes, properties
of resonances like W, Z, Higgs boson or in searches for New Physics.

That is why, it was of interest to prepare a package called  {\tt TauSpinner} which 
can be used for correcting with weights distinct effects in generated event
samples including $\tau$ decays. {\ZW In many cases one can argue that such
re-weighting of sample with different physics assumptions may be of little
use. Why not simply generate another series of events with new physics 
assumptions and compare obtained from them distributions? Indeed one can
think re-weighting may bring little benefits and considerable risks, eg. if
the original sample, has some sectors of the phase space biased by a  
particular technical cut-offs. Or if it features zero
matrix element on some hypersurface within the phase space manifold. Then
arbitrary large weights may start to appear.
 However, in several cases such disadvantages can be balanced by benefits.
The weighted events sample is statistically correlated with the original one,
thus statistical errors affect the estimate of the differences only. 
Even larger CPU gain is expected, if the sample is processed with
simulation of the detector effects. 
The detector effect simulation usually requires few orders more CPU time 
than generation of physics events, or calculation of events weight.
Even the inclusion of Next to Leading Order (NLO) effects in the generation,
can be CPU costly as well. Reweighting to different production process with the
same final state (set of final states) can be of interest.
That is why the reweighting algorithms  are prepared; simultaneously
%
of $\tau$-pair production and $\tau$ decays. 
The weigt may depend on kinematical configurations consisting of ten or even more  four momenta. 
}

 At the beginning of the project the interest was focused on longitudinal 
spin effects only \cite{Czyczula:2012ny}. The   {\tt TauSpinner} 
was used to calculate appropriate correcting weight.
In this way such spin effects could be included (or removed) 
into the generated events sample. With time, other options  
were added. Since  \cite{Banerjee:2012ez}, additional weight was introduced, 
to manipulate production process adding  or replacing generated production process with
the alternative one including for example exchange of new intermediate 
particles. Publication \cite{Kaczmarska:2014eoa} brought possibility of introducing complete spin 
effects (longitudinal and transverse) in decay of intermediate Higgs, and  \cite{Przedzinski:2014pla} in case of Drell-Yan process as well.
Production process was modelled with lowest order $2 \to 2$ matrix elements.
In  \cite{Kalinowski:2016qcd} implementation of hard processes featuring parton level 
matrix elements for production of $\tau$ lepton pair and two jets was introduced.
This was motivated by the experimental analyses becoming sensitive to Electroweak Higgs and $Z$ boson production in fusion of $WW$ or $ZZ$ pair. The question of
the systematic uncertainties due to approximation where $2 \to 2$ or $2 \to 4$
matrix elements for calculation of spin effects could be adressed since then.
With time also more sophisticated options for treatment of spin effects was prepared. 
In \cite{Kaczmarska:2014eoa} possibility to attribute helicity to $\tau$-leptons was added.
It uses approximation, only longitudinal components of density matrix for 
$\tau$ pair production was taken into account.
In \cite{Przedzinski:2014pla} one loop electroweak corrections in case of Drell-Yan process became available for the weight calculation.   
Finally let us mention, that available options allow to configure {\tt TauSpinner} algorithm to work 
on  samples where only partial spin effects were taken into account and to correct them to 
full spin effects.

All these different options were introduced into {\tt TauSpinner} because of user's 
interest and under pressure of time during converging very fast analyses of LHC Run I data by ATLAS and 
CMS experiments. That is why, there is a need and opportunity now to document 
them systematically 
and archive the code which will be used during the final  LHC Run II analyses.
 
Contrary to previous papers, we will emphasise here less phenomenological applications
but rather focus on common 
theoretical/phenomenological basis and relations between different options.
Particularly, we will bring the point of the theoretical uncertainties of the 
proposed algorithms only briefly. 
For quantified results we will refer the reader to previous publications.

Our paper is organized as follows. In Section~\ref{sec:Basis} theoretical basis
for describing $\tau$'s in production and decay matrix elements,
is introduced. We define weights calculated by the program.
Details of   kinematics; definition of frames  and connecting them transformations 
used for calculation of hard-parton level amplitudes or cross section and 
for calculation of spin effects  are  given
in Sections~\ref{sec:Wtcal} and~\ref{sec:ExaAppr}.
In  sub-sections 
we explain  simplifications and resulting formulae for the differential 
cross section and spin correlations.
Section~\ref{sec:TwoToTwo} is devoted to the case when at parton level $2 \to 2$ processes 
are used.    
Section~\ref{sec:HHcalc} covers calculation of density matrices for $\tau$ decays; the
polarimetric vectors. In Section~\ref{sec:VBF} the case of  $2 \to 4$ processes is discussed and 
explained are extensions to the algorithms which were needed for the case when final state
consists not only of $\tau$ lepton pair, but also two additional jets are present.
Appendices recall all  examples and
applications which are included in the distribution. So far, documentation for
these applications was scattered  over several publications appendices and 
thus was not available in a systematic manner.
Summary~\ref{sec:Summary}, closes the paper.

\section{Theoretical basis} \label{sec:Basis}

We start discussion from presenting exact formula on the cross section for 
$pp \to \tau \tau\; X$ or  $pp \to \tau \; \nu_\tau\; X$ processes, 
which requires knowledge of 
the whole process matrix elements, including decays of $\tau$ leptons.
With about 20 $\tau$ decay channels we end-up with  400 possible 
configurations for each $\tau$-pair production process.
The exact  formula can not be therfore used in practice.
Nonetheless, it can serve as a starting point for the more practical
to use ones, also those 
involving  approximations. The formulae presented below 
for the $ H (Z/\gamma^*) \to \tau \tau $
can be applied also to the
 $H^\pm  (W^\pm) \to \tau \; \nu_\tau $ case.
Then the second $\tau$ of the formulae should be replaced by $\nu_\tau$.
Consequently, part of formula which describes its decay is dropped out.

\subsection{Exact formula} \label{sec:BasisExact}

The basic formula for 
cross-section for the process $ pp \to \tau^+\tau^- X ;  \tau^+ \to Y^+ \bar \nu; \tau^-\to Y'^- \nu$ reads:

\begin{equation}
d \sigma = |{\cal M}|^2 d\Omega= |{\cal M}|^2 d\Omega_{prod} \; d\Omega_{\tau^+} \; d\Omega_{\tau^-} \label{eq:basic}
\end{equation}
Where ${\cal M}$ denotes complete matrix element.
Obvious terms due to average over initial state spin degrees of freedom  
and flux factor are dropped out. Because of the $\tau$ lepton width being very 
small, the $\tau$-lepton propagator module squared, to a perfect approximation,
 reduces to the Dirac $\delta$ function in lepton virtuality and so the 
 infinitesimal element in the kinematical phase-space factorizes into $d\Omega_{prod} \; d\Omega_{\tau^+} \; d\Omega_{\tau^-}$.

The formula (\ref{eq:basic}) is not very convenient,  because of over 20 possible $\tau$ decay channels which lead to 400 distinct processes
and matrix elements. Fortunately, as for the phase space parametrization, because of a narrow $\tau$ width,
 matrix element can be also expressed  as product  of amplitudes for 
production and decays, summed over each $\tau$ spin projections ($\lambda=1,2$) on at this point unspecified axes

\begin{equation}
{\cal M}=\sum_{\lambda_1 \; \lambda_2=1}^2{\cal M}_{\lambda_1 \; \lambda_2}^{prod} \; 
      {\cal M}_{\lambda_1}^{\tau^+}{\cal M}_{\lambda_2}^{\tau^-}.
\end{equation}

The cross-section can be expressed then with the rewritten 
exact formula,  which  thanks to Fierz identity decouples summations of the decay and production
matrix elements. It introduces the correlations back in a form of spin weight,

\begin{equation}
d \sigma = \Bigl(\sum_{\lambda_1\lambda_2 }|{\cal M}^{prod}|^2 \Bigr)
 \Bigl(\sum_{ \lambda_1}|{\cal M}^{\tau^+}|^2 \Bigr)
 \Bigl(\sum_{\lambda_2}|{\cal M}^{\tau^-}|^2 \Bigr) wt_{spin} \;
 d\Omega_{prod} \; d\Omega_{\tau^+} \; d\Omega_{\tau^-}. \label{eq:useful}
\end{equation}
This formula is 
a core of {\tt TauSpinner} algorithm, but not only. It is  used in
{\tt Tauola}
$\tau$ decay library interface as
foundation of the ``after-burn'' approach~\cite{Davidson:2010rw} and also in {\tt KKMC} Monte Carlo
\cite{Jadach:1999vf}.
For the details of geometrical formulation and naming conventions
Ref.~\cite{Jadach:1998wp} may be useful.

The spin weight $wt_{spin}$ is dimensionless,  and contains information of all spin effects transmitted 
from the production to the decay of $\tau$ leptons,

\begin{equation}
wt_{spin} = \sum_{i,j=t,x,y,z} R_{i,j} h^i_{\tau^+} h^j_{\tau^-}. \label{eq:wtspin}
\end{equation}
The indices $i,j$,  of the adjoint (to 1/2 representations of fermions)
rotation group representation run over $t,x,y,z$.  
The polarimetric vector $h^i_{\tau^\pm}$ is calculated from formula 
\begin{equation}
h^i_{\tau^{\pm}} =\sum_{\lambda, \bar \lambda} \sigma^i_{\lambda, \bar \lambda}  {\cal M}_{\lambda }^{\tau^{\pm}} {\cal M}^{\tau^{\pm} \; \dagger}_{\bar \lambda }, \label{eq:hi}
\end{equation}
where  $\sigma^i_{\lambda, \bar \lambda}$ stands for Pauli matrices. The $ h^i_{\tau^{\pm}}$ are normalised further to set their time-like component to 1, 
namely taking $h^i_{\tau^{\pm}}=\frac{h^i_{\tau^{\pm}}}{h^t_{\tau^{\pm}}}$.
We will often use shortened form for the notation  $h^i_{\tau}$ or  $h^i_{\pm}$. The superscript
$i$ will be dropped if we will have all components in mind.

The spin correlation matrix $R_{i,j}$ is calculated from formula
\begin{equation}
R_{i,j}=\sum_{\lambda_1, \bar \lambda_1 \lambda_2, \bar \lambda_2}\sigma^i_{\lambda_1, \bar \lambda_1}  \sigma^j_{\lambda_2, \bar \lambda_2}  
      {\cal M}^{prod}_{\lambda_1 \lambda_2 } {\cal M}^{prod \; \dagger}_{\bar \lambda_1 \bar \lambda_2 } \label{eq:Rij}
\end{equation}
and also normalized to set its time-like component to 1, namely $R_{i,j}=\frac{R_{i,j}}{R_{t,t}}$.

Formula (\ref{eq:wtspin}) is universal, it does not depend on the production process,
and there are no approximations  introduced. 
It features useful properties: $0 <wt_{spin}<4$  and weight average  $<wt_{spin}>=1$.
Use of expressions  (\ref{eq:hi}) and (\ref{eq:Rij})  moves our description 
from language of spin amplitudes toward 
the language of cross-sections and probability distributions; the amplitudes squared contribute
only.

All complexities are in calculation of $R_{i,j}$ spin correlation matrix for production and  $h^i_{\tau^+} h^j_{\tau^-}$  
polarimetric vectors for the decay of $\tau$ leptons.
The $R_{i,j}$ matrix describe the full spin correlation between the two $\tau$'s as well as the individual 
spin states of the $\tau$'s. 
The $R_{i,j}$ depends on kinematics of production process only, 
and $h^i_{\tau^+} h^j_{\tau^-}$ respectively on kinematics of $\tau^+$ and $\tau^-$.
The explicit definition of the  matrix $R_{i,j}$ and vectors   $h^i_{\tau^+}$, $h^j_{\tau^-}$ 
is rather lengthy and also well known, see~\cite{jadach-was:1984,Jadach:1993hs}
for detailed definitions. It is decay channel dependent.

So far we discussed case of  $\tau$ pair in the final state, that is the case of $H \to \tau \tau$ or $Z/\gamma^* \to \tau \tau$ processes.
Case of only one $\tau$ in final state, that is the case of $W^\pm\to \tau \; \nu_\tau$ or $H^\pm\to \tau \; \nu_\tau$ is much simpler. The $R_{i,j}$ matrix is replaced by a vector 
of components $R_t,R_x,R_y,R_z= 1,0,0,\mp 1$ respectively for $W$ and charged Higgs decays.
Sum over two indices is thus reduced to sum over one index only, axis $z$ is the direction of $\tau_\nu$ from 
$W^\pm/H^\pm$ decay as seen from $\tau$-lepton rest-frame, spin weight read thus as $wt_{spin}= 1\pm h_{\tau}^z$. 
The frames choices (presented in the next Section) for that case
could have been simpler than  for $Z/H$ decays, no details of transverse directions 
are then needed.

\subsection{Formula using parton level amplitudes} \label{sec:partonlevel}
We do not have at our disposal matrix elements for the complete  $pp \to \tau^+\tau^- X$ process, 
but instead, following factorisation theorems, we can represent cross-section 
with the help of the parton level amplitudes,  convoluted with the parton 
density functions (PDF's) and summed over all $flavours$ (flavour configurations of 
incoming partons\footnote{The double sum, runs in principle
  over all possible flavours  for the two incoming partons:
  $g, u, \bar u, d, \bar d, c, \bar c, s, \bar s, b, \bar b$.
}. Thus, from  formula (\ref{eq:useful}) we obtain:

 \begin{eqnarray}
&d \sigma = \sum_{flavours} \int dx_1dx_2 f(x_1,...)f(x_2,...)d\Omega^{parton\; level}_{prod} \; d\Omega_{\tau^+} \; d\Omega_{\tau^-} \nonumber \\
&\Bigl(\sum_{\lambda_1,  \lambda_2 }|{\cal M}^{prod}_{parton\; level}|^2 \Bigr)
 \Bigl(\sum_{\lambda_1 }|{\cal M}^{\tau^+}|^2 \Bigr)
 \Bigl(\sum_{\lambda_2 }|{\cal M}^{\tau^-}|^2 \Bigr) wt_{spin}. \label{eq:parton-level}
\end{eqnarray}

 At this point we would like to decouple calculation of $wt_{spin}$ as much as
 possible from the  PDF's dependence. Note,
that this is needed for the $Z/\gamma^*$ mediated processes due to $V-A$ structure of 
$Z$ boson and $\gamma^*$ couplings to fermions. For the $H$ or $W$ mediated processes spin polarization
state does not depend on flavours of incoming partons. For the $Z/\gamma^*$
the $R_{i,j}$ used in calculation of $wt_{spin}$, is taken as weighted average
(with PDFs and production matrix elements squared) 
over all flavour configurations, as in the following equation:

 \begin{eqnarray}
R_{i,j} \to \frac{ \sum_{flavours} \;\Bigl( \;f(x_1,...)f(x_2,...) 
\Bigl(\sum_{\lambda_1,  \lambda_2 }|{\cal M}^{prod}_{parton\; level}|^2 \Bigr) \; R_{i,j}\;\Bigr) }
{\sum_{flavours}\;\Bigl(\;  f(x_1,...)f(x_2,...) 
\Bigl(\sum_{\lambda_1,  \lambda_2 }|{\cal M}^{prod}_{parton\; level}|^2 \Bigr)\;\;\;\;\;\;\;\;\;\Bigr)} . \label{eq:Rij-ave}
\end{eqnarray}
No  approximation is introduced in this  way,
denominator of Eq. (\ref{eq:Rij-ave}) cancels explicitly the corresponding
factor of Eq.~(\ref{eq:parton-level}) (that is also the integrand of 
the following Eq.~(\ref{eq:trouble}), first line).

 To introduce corrections due to different spin effects, modified production
 process or decay model,  in the generated sample (i.e. without re-generation
 of events)  one can define the weight $WT$, representing
 a ratio of new to old cross-sections at each point in the phase-space.

The Eq.~(\ref{eq:parton-level}) for modified cross-section, takes then the form
 \begin{eqnarray}
d \sigma &=& \sum_{flavours} \int dx_1dx_2 \;\; f(x_1,...)f(x_2,...)d\Omega^{parton\; level}_{prod}  
\Bigl(\sum_{\lambda_1,  \lambda_2 }|{\cal M}^{prod}_{parton\; level}|^2 \Bigr) \nonumber \\
&& \Bigl(\sum_{\lambda_1 }|{\cal M}^{\tau^+}|^2 \Bigr) \; d\Omega_{\tau^+}\; \; \; 
 \Bigl(\sum_{\lambda_2 }|{\cal M}^{\tau^-}|^2 \Bigr)  \; d\Omega_{\tau^-}  \label{eq:trouble}   \label{eq:global}  \\ 
&& \times \; wt_{spin} \;  \times \; WT. \nonumber 
\end{eqnarray}
 Variant  of Eq.~(\ref{eq:global}), described in the footnote%
 \footnote{It would seem natural to make a choice of 
the initial flavours, on the basis of probabilities obtained from production density 
$1/N_{normalization}\; f(x_1,...)f(x_2,...)\bigl(\sum_{\lambda_1,  \lambda_2 }|{\cal M}^{prod}_{parton\; level}|^2 \bigr)$ over the flavours.
This would also seem quite similar to what is used in 
{\tt Tauola interface} \cite{Davidson:2010rw}. Such solution would not be correct. The dependence on the decay configuration,  
already fixed and stored for the event, would be then ignored.
In case of {\tt Tauola interface} this information can be obtained from
the event record.},
 is used in {\tt Tauola}.
 
In general the  weight $WT$, 
factorises into multiplicative components/weights: production ($wt_{prod}$), decay 
($wt_{{decay}}^{\tau^\pm}$) and ratio of spin correlation/polarization weights, new to old one 
${wt_{spin\; new}}/{wt_{spin\; old}}$: 

\begin{equation}
WT = wt_{prod} \; \; wt_{decay}^{\tau^+} wt_{decay}^{\tau^-}{wt_{spin\; new}}/{wt_{spin\; old}}.\label{eq:combi}
\end{equation}
The first three terms of the weight, represent modification of the matrix elements for production and decays 
\begin{eqnarray}
wt_{prod}&=&  \frac{\sum_{flavours}f(x_1,...)f(x_2,...)\Bigl(\sum_{spin }|{\cal M}^{prod}_{parton\; level}|^2\Bigr) \Big|_{new}}
                {\sum_{flavours}f(x_1,...)f(x_2,...)\Bigl(\sum_{spin }|{\cal M}^{prod}_{parton\; level}|^2\Bigr) \Big|_{old}},\label{eq:prod}\\
 wt_{decay}^{\tau^\pm}&=&\frac{\sum_{spin }|{\cal M}^{\tau^\pm}_{new}|^2}{\sum_{spin }|{\cal M}^{\tau^\pm}_{old}|^2}. \label{eq:dec}
\end{eqnarray}
The  combined weight as in Eq. (\ref{eq:combi}), is nothing else than ratios of spin averaged amplitudes squared 
for the whole process; new to old. The eventual changes in the PDFs parametrizations should be taken into account 
in calculation of $wt_{prod}$ and also in ratio ${wt_{spin\; new}}/{wt_{spin\; old}}$
in formula (\ref{eq:combi}). In case production process or decay models are not modified, respectively  
$wt_{prod}$ and $wt_{decay}^{\tau^{\pm}}$ are equal to 1.
Ratio of spin weights $wt_{spin\; new}/wt_{spin\; old}$ allows for introduction of new spin effects.  
In case of originally  unpolarized sample, the  $wt_{spin\; old} = 1$ and   $wt_{spin\; new}$ alone allows to introduce desired spin effects.

Formula (\ref{eq:trouble}) can be rewritten to equivalent form, which  underlines separation useful for  the organization of program, with
segments 
for production of $\tau$ leptons, their  
decays and   the spin weights. Let us recall that the spin weight $wt_{spin}$ 
is the only part depending simultaneously on production and decay kinematics
and for production of $\tau$ lepton pair 
has always a form of Eq.~(\ref{eq:wtspin}). Ratio of this weight 
calculated from new and old matrix elements has then to be taken. Reorganizing
some terms we can finally write:

\begin{equation}
d\sigma = \Bigl[ d\sigma_{prod}\;\; d\Gamma_{decay}^{\tau^+} d\Gamma_{decay}^{\tau^-}\Bigr]\; wt_{spin\; old}\;\;  \Bigl(\frac{wt_{spin\; new}}{wt_{spin\; old}}\Bigr),
\end{equation}
where

\begin{eqnarray}
d\sigma_{prod}&=& \sum_{flavours} \int  dx_1dx_2 f(x_1,...)f(x_2,...)d\Omega^{parton\; level}_{prod} \Bigl(\sum_{spin }|{\cal M}^{prod}_{parton\; level}|^2 \Bigr) wt_{prod} \nonumber \\
d\Gamma_{decay}^{\tau^+} &=&   d\Omega_{\tau^+} \Bigl(\sum_{spin }|{\cal M}^{\tau^+}|^2 \Bigr)  wt_{decay}^{\tau^+},\\
d\Gamma_{decay}^{\tau^-} &=&  d\Omega_{\tau^-} \Bigl(\sum_{spin }|{\cal M}^{\tau^-}|^2 \Bigr)wt_{decay}^{\tau^-}.\nonumber
\end{eqnarray}
All four component of the overall weight (\ref{eq:combi}) are given separately
to underline, wherever possible, their dependence on all or only sub-set of
phase-space dimensions: production and/or decay.
Note that for $W^\pm/H^\pm \to \tau\; \nu_\tau$ we have
not prepared so far any option which allows to modify production processes.

\section{WT calculations in TauSpinner}\label{sec:Wtcal}

The formulae listed above are used in {\tt TauSpinner} for calculation of distinct 
components of $WT$ weight, see Eq. (\ref{eq:combi}). 
Let us start with the main weight of the spin effects, $wt_{spin}$ defined by 
Eq.~(\ref{eq:wtspin}). With this weight, 
for samples where  spin effects of $\tau$ production are absent, they can be
inserted into decays distributions. Alternatively, its inverse  can be used
to remove spin effects from the sample where
they are taken into account. 
A building block of this weight introducing the dependence on the production 
process, the matrix $R_{i,j}$, is given  by the formula (\ref{eq:Rij-ave}).
The other weights,  $w_{prod}$ and $w_{decay}^{\tau^\pm}$ are defined by formulae (\ref{eq:prod}, \ref{eq:dec}).

 It is useful to 
introduce  notation  $P^z_{\tau}=R_{t,z}=R_{z,t}$, which represents 
longitudinal polarization of the single $\tau$ 
(if one integrates out possible configurations of the other $\tau$).
In  usually sufficient approximation of helicity states, that is when 
transverse momenta of $\tau$ decay products are neglected,
as is  the case of ultrarelativistic $\tau$ leptons, the  $P^z_{\tau}$ is the 
only non-trivial (dynamic dependent) element of matrix $R$. All others are equal $\pm$1, 
or can be set to 0.
We will return to the details later.

In calculation of $wt_{spin}$ and $w_{prod}$ physics uncertainty depend on
accuratness of factorization assumptions for separation of patron level
matrix elements and PDFs; also on  choice of particular PDFs parametrization. 
Nothing of principle would change if instead of $2\to 2$ body production matrix
element one would use   $2\to 2+n$ where $n$ denotes additional partons/jets 
or other particles (of summed spin-states). Except the choice of PDFs
(eventually also model of underlying event interactions), for a parton level
matrix elements one has to make a careful 
choice of how the hard scattering kinematics is reconstructed from information
available in the event. 

Calculation of $wt_{decay}^{\tau^{\pm}}$ involves modelling of $\tau$ decays only. It is of interest for studies of the resulting
systematic errors. This part of the code is covered by the {\tt Tauola} library.

We will return to these points later, but one has to keep it in mind already now, while reading the following.

\subsection{Kinematical frames} 
\label{sec:frames}

Components necessary for  the $WT$ weight are calculated in different frames. This is a correct approach as long as
details of boosts and rotations  connecting frames are meticulously followed. Also spin states of the $\tau$ leptons may
be defined in different frames than four momenta of the hard process.
Bremsstrahlung photons and properties
of the  matrix element in their presence, require dedicated treatment. 

Basic formula (\ref{eq:basic}) can in principle provide exact results. In practice, we have 
to introduce approximations to adapt to the fact that matrix elements are 
calculated from the parton level amplitudes, and  to the way how these amplitudes
are reconstructed by {\tt TauSpinner} algorithm where only four momenta of outgoing
$\tau$ leptons and partons can be used. Definition of quantization frames is an essential element of 
the arrangements, let us provide now the necessary details.

For calculating all components of event weights (to prepare kinematical configurations of the events
necessary for that purpose) we define in total four frames (group of frames):
\begin{enumerate}
\item[$A:$] Basic frame (starting point) for kinematic transformations and all other frames definitions.
  The rest frame of $\tau$-pair is used (and not
  of $\tau$-pair with the final state bremsstrahlung photons combined).
  Such a choice is possible thanks to 
  properties 
of bremsstrahlung amplitudes. Emitted photons do not carry out the spin.
This non-trivial observation is exploited also in {\tt Photos} Monte Carlo phase space 
parametrization, see Refs.~\cite{Davidson:2010ew,Nanava:2009vg}.
The definition of the rest frame of the $\tau$ lepton pairs, is completed with incoming partons set along the $z$-axis.
\item[$A':$]
The rest frame of the $\tau$ lepton pairs and final state bremsstrahlung photons included. Incoming partons are again set along the $z$-axis. This frame is 
used for calculating production weight $wt_{prod}$ and spin correlation matrix $R_{i,j}$; if no bremsstrahlung photons are present it is
equivalent\footnote{ In case when photons are present and for $2 \to 4$ matrix elements
  modifications are needed to assure that on-mass shell kinematic configuration is passed to the routines calculating spin amplitudes.} to
frame A.  In every case we reconstruct $x_1, x_2$, the arguments of
PDF function, 
from virtuality $M$ and longitudinal to the beam direction component ($p_L$) of the intermediate state (sum of momenta of 
$\tau^+\; \tau^-\; n\gamma$) momentum in lab frame. The center of mass energy of $pp$
scattering is used to calculate $x_1 x_2 $ through relations
$x_1 x_2 CMS_{ENE}^2 = M^2(\tau^+\; \tau^-\; n\gamma)$,
 $(x_1-x_2) CMS_{ENE} = 2 p_L(\tau^+\; \tau^-\; n\gamma)$. The $n\gamma$ is assumed to correspond to final state bremsstrahlung
associated with $\tau$-pair production and its momentum has to be taken into account, but only in calculation of virtuality
for intermediate $Z/\gamma^*$ state { in case} of $2\to 2$ amplitudes\footnote{
  In case when configurations with two-jets are evaluated ($2\to 4$ processes),
  not only more attention for the bremsstrahlung
  photons will be needed, but also the momenta of the jets have to be taken into account in evaluation of $p_L$ and $M^2$. We will
return to this point in the future.}. Note that $A'$ is used for  $x_1, x_2 $ calculation only. The $2\to 2$ hard process scattering angle is calculated in frame $A$.
\item[$B:$] The
rest frame of the $\tau$ lepton pairs, with $\tau$-leptons along the z-axis.
It is used for common orientation of production and decay coordinate systems, in this frame
defined are polarimetric vectors $h_{\pm}$. Transformation $A \rightarrow B $ requires rotation.
\item[$C_{\tau^+}, C_{\tau^-}:$] The
rest frames of individual $\tau$ leptons, with direction of the boost 
to $\tau$ pair rest-frame along the z-axis. 
Transformation $B \rightarrow C_{\tau^+}$ or $C_{\tau^-}$ 
requires boost only.
\item[$D_{\tau^+}, D_{\tau^-}:$] The
rest frames of individual $\tau$ leptons, with $\nu_{\tau}$ along the z-axis. 
It is used for calculating decay weights. In fact in the frames $D_{\tau^{\pm}}$ also  polarimetric 
vectors $h_{\tau^{\pm}}$ are calculated and then rotated back to frame $C_{\tau^{\pm}}$. 
 
\end{enumerate}

In Table \ref{Tab:FRames}, we summarize which frames are used for calculations
of different variables and then to which frames they are boosted (rotated) before
being used in formula (\ref{eq:parton-level}). Let us stress, that it is not
only important to define all frames of type $A-D$ but also Lorentz transformations 
between them.

The frame of type $A$ is used for calculation of hard process amplitudes.
If only longitudinal components of spin density matrix are taken into account,  
we do not need to define in detail transverse directions (quantization frame versors) with respect of $\tau$ momenta, 
and control the corresponding details of boost methods to the rest frames of $\tau$'s 
is not needed. However, the code of {\tt TauSpinner}  controls such details 
of the boosts, because it is prepared to handle also cases when complete 
spin density matrix is used.

Frame of type B and frames of type C have common direction of the $z$-axis, 
also $x$- and $y$- axes coincide. Frame B is used for $R_{i,j}$ calculation if transverse degree of freedom are taken into account.
Calculation of the spin weight, 
that is contraction of indices in
$wt_{spin} =  \sum_{ij} R_{i,j} h^i_{\tau^+} h^j_{\tau^-}$, is performed in frames
$C_{\tau^+}$, $C_{\tau^-}$. It enable control  on transverse spin effects. 
Frames of type $D$  are used 
for calculation of decay matrix elements and $h^{i}_{\tau^+},h^{i}_{\tau^-} $
polarimetric vectors with methods from {\tt Tauola} library,
see Ref.\cite{Jadach:1993hs} 
for  details.

 Let us note, that
we have used such frames in many applications  already at a time of LEP  analyzes, 
they were useful for comparisons and tests. These choices once established, can be easily modified, whenever necessary,
 e.g. to fulfil constraints of particular conventions of spin amplitudes calculations, see 
eg. in Ref.~\cite{Jadach:1998wp}.
With the definition of frames completed, we can continue our discussion.

\begin{table}
\vspace{2mm}
\begin{center}                               
\begin{tabular}{|l|r|r|r|}
\hline \hline
Object                     & Frame of object calculation & Frame of object use & Comment                         \\
\hline 
$ R_{i,j}  $ for  $wt_{spin}$ & $A$ (or $B$)     & $C^\pm$     & $B$ if tranverse                \\
                           &                  &            &     spin included               \\
$h^i_{\tau^+}, h^j_{\tau^-} $ for  $wt_{spin}$ &  $D^\pm$         &  $C^\pm$    &                                 \\
$wt_{prod} $                &  $A$ ($A'$)      &            &   $A'$ is used if brems-        \\
                           &                  &            &   strahlung photons are present  \\
$wt_{decay}$                &  $D^\pm$         & $D^\pm$     &                                  \\
\hline 
\end{tabular}
\end{center}                               
\caption{ Frames used for calculation of components used in weights
 calculation,  Eq.~(\ref{eq:combi}).
 Specified are also frames to which these components need to be transformed.  
 \label{Tab:FRames} }

\end{table}

{\ZW
  \subsubsection{Variants for definition of $\tau$ lepton pair rest frame.}
Let us comment on the possible future improvement for the definition of the
frame $A$, potentially useful freedom of choice is available.
At present, in {\tt TauSpinner}  definiton of lepton pair rest-frame {\it A},
beam direction, necessary to define Born-level scattering angle,
the simplest possible ansatz, 
that in a given event there are no jets of substantial $p_T$, is used. This can be improved without 
a need to re-do the design of the presented above tree of frames definitions. 
The necessary studies for matrix elements featuring one or two high $p_T$ 
jets are already documented in Refs.~\cite{Richter-Was:2016mal,Richter-Was:2016avq}.
The corresponding modifications of the code were not yet introduced
into {\tt TauSpinner}, as there are no clear indications of their numerical
importance, but it is an input for the possible forthcoming improvement.
}

\subsection{Production weight $wt_{prod}$ and spin correlation matrix $R_{i,j}$ }

The  $R_{i,j}$ matrix is not a Lorentz invariant tensor. In fact, its indices run over coordinates in {\it two}  frames,   $C_{\tau^{+}}$
and $C_{\tau^{-}}$ at the same time as the input for matrix element calculation the  four momenta of  $A$ frame are used. For details of
relations between frames and elements used in weights calculation, see also Table \ref{Tab:FRames}.

For calculation of production weight $wt_{prod}$ and  spin correlation matrix $R_{i,j}$,
centre-of-mass frame of the $\tau$ lepton pair with incoming partons being along z-axis (frame of type $A$) is not the optimal one.
It is nonetheless used if there is no interest in transverse spin degrees of freedom.
Frame of type $B$,
with $\tau$-leptons  along $z$-axis, is more convenient and with $x$ and $y$ axes parallel to the ones of frames $C_{\tau^{\pm}}$
completes the  most convenient setup.
In fact all these three frames $B$ and   $C_{\tau^{\pm}}$ are used {\it simultaneously}
for calculation of spin weight of Eq.~(\ref{eq:wtspin}).

The kinematical frames for $\tau$-pair production and $\tau$ decays need to be explicitly
related by the set of Lorentz transformations with the frames of matrix elements calculations.
Let us recall details in a form of algorithmic steps, which need to be performed
on all decay products of $\tau$ leptons before calculation of $R_{i,j}$
and $h^{i/j}_{\tau^\pm}$ (for $h^{i/j}_{\tau^\pm}$ further steps will be still needed): 
\begin{itemize}
\item
Boost both $\tau$ leptons and their decay products from the
laboratory frame  to rest-frame of $\tau \tau$ pair. 
\item
Rotate all to the frame, where direction of incoming partons is along the $z$-axis (frame  $A$).
\item
  Rotate so that $\tau$ leptons are set along the z-axis and incoming partons remain in $z-y$ plane  (frames $C_{\tau^\pm}$ have to be correlated with  $B$ by a boost along $z$ axis).
  These frames are used for  $R_{i,j}$ spin states definition too.
\item[-]
Presence of the final state bremstrahlung from $\tau$ pair brings complication, because the rest frame of the 
$\tau$ pair is not the rest frame of the resonance which decayed into $\tau$ pair.
For the sake of kinematical transformations, frames $A$ and $B$ defined as above are used. 
However, for calculation of matrix elements we  use frame $A'$, where efffectively 
bremsstrahlung photons are absorbed into $\tau$ lepton momenta.
For calculation of $x_1, x_2$ we use invariant mass, which can be easily calculated in frame $A'$.
\end{itemize}

\subsection{Decay weight $wt_{decay}^{\tau^{\pm}}$ and polarimetric vectors $h_{\tau^{\pm}}$ }

The decay weight $wt_{decay}^{\tau^{\pm}}$  and polarimetric vectors $h_{\tau^{\pm}}$ are calculated for each outgoing $\tau$ lepton
in its rest frame, denoted as frame $D_{\tau^{\pm}}$ that is a frame where neutrino from $\tau$ lepton decay, $\nu_{\tau}$ 
is along z-axis. The Lorentz transformations of the previous sub-section have to be supplemented with the one relating frames  $C_{\tau^{\pm}}$ with $D_{\tau^{\pm}}$.

\begin{itemize}
\item
Boost all decay products along z-axis to the rest frame of the $\tau^{\pm}$ lepton
(frames  $C_{\tau^{\pm}}$).
\item
Rotate again $\tau$ daughters so that $\tau$ neutrino is along z-axis (frames  $D_{\tau^{\pm}}$).
\item
Calculate in frames  $D_{\tau^{\pm}}$ polarimetic vectors $h^i_{\tau^\pm}$,
 rotate them  back
 from $D_{\tau^{\pm}}$ to $C_{\tau^{\pm}}$. Only then, they can be contracted with $R_{i,j}$ for $wt_{spin}$ 
calculation.

\end{itemize}

Decay weights and polarimetic vectors for $\tau^+$ and $\tau^-$ are calculated in frame type $D_{\tau^{\pm}}$,
using code from {\tt Tauola}, which sets  requirement of $\nu_\tau$  along z-axis.

\section{Exact and approximate spin weight $wt_{spin}$ }\label{sec:ExaAppr}
Basic formula (\ref{eq:useful}) of Section  \ref{sec:BasisExact} is exact. Because of approximated 
calculations (introduction of parton level amplitudes) or as explained later in the Section, 
to obtain easier to interpret physical picture 
(valid in ultrarelativistic case only) we can introduce simplifications. 
 No such simplification was introduced yet.
Let us recall first that according to $R_{i,j}$ and $h_{\tau^{\pm}}^i$ definitions, $R_{t,t}=1$ and
  $h_{{\pm}}^t=1$ (we can use shorter notation $h_{{\pm}}^i=h_{\tau^{\pm}}^i$).

\subsection{ Neglecting $m_\tau^2$ terms}
With the choices of  kinematical frames as defined 
in Section \ref{sec:frames} and neglecting   $m_\tau^2$ terms following the  ultrarelativistic limit,   
one gets that $R_{z,z} = sign = \pm 1$
(the $sign$ depends whether decaying object is a scalar or a vector boson). 
Also, non-diagonal terms
$R_{t,x}=0$, $R_{t,y}=0$, $R_{x,t}=0$, $R_{y,t}=0$, $R_{z,x}=0$, $R_{z,y}=0$, $R_{x,z}=0$, $R_{y,z}=0$,
independently whether the decay of spin zero (scalar) or spin one (vector) state is considered.
The formula (\ref{eq:wtspin}) reduces to:

\begin{equation}
wt_{spin} =  1+ { sign} \ \ h^z_{\tau^+} h^z_{\tau^-}
             + R_{z,t} h^z_{\tau^+}  + R_{t,z} h^z_{\tau^-}
             + \sum_{i,j=x,y} R_{i,j} h^i_{\tau^+} h^j_{\tau^-}, \label{eq:wtspinner0}
\end{equation}
or if notation of individual  $\tau$ longitudinal polarization $P^z_{\tau}$ is used (for Higgs decays $P^z_{\tau}=0$),
\begin{equation}
wt_{spin} =  1+ { sign} \ \ h^z_{\tau^+} h^z_{\tau^-}
             + P^z_{\tau} h^z_{\tau^+}  + P^z_{\tau} h^z_{\tau^-}
             + \sum_{i,j=x,y} R_{i,j} h^i_{\tau^+} h^j_{\tau^-}. \label{eq:wtspinner1}
\end{equation}
Let us stress, that the formulae (\ref{eq:wtspinner0}, \ref{eq:wtspinner1}) are quantization frame dependent. One has to 
be careful especially with signs, if the kinematical transformations,
 as explained in Section
\ref{sec:frames} are not followed. Distinct conventions for defining those frames 
are used in the literature.


\subsection{ Neglecting transverse spin correlations and spin state probabilities}\label{sec:negl}

Further approximation (not always used) is  to include 
longitudinal spin effects only, means that terms  $R_{i,j}$  are set to zero for 
$i,j=x,y$. In general, those terms can be large, but as they result in dependencies
of transverse with respect to $\tau$ direction components of $\tau$ decay products
momenta, often do not lead to measurable effects and can be dropped.   

With this approximation, formula (\ref{eq:wtspinner1}) becomes
\begin{equation}
wt_{spin} =  1+ { sign}\ h^z_{+} h^z_{-}
             + P^z_{\tau} h^z_{\tau^+}  + P^z_{\tau} h^z_{\tau^-}, \label{eq:wtspinner2}
\end{equation}

Using formula (\ref{eq:wtspinner2}), and resulting approximation, one will be able to introduce  probabilities for spin states $p_\tau^z$ which will be defined later.
For decay of scalar resonance (eg. Higgs) we define (equal) probabilities for $ \tau $ 
pair configurations being left-right ($+$) or right-left ($-$). We  rewrite formula (\ref{eq:wtspinner2})
and decompose $wt_{spin}$ into sum of $wt_{spin}^+$ and  $wt_{spin}^-$:
\begin{equation}
wt_{spin} =  0.5 (1+  h^z_{\tau^+}) (1- h^z_{\tau^-})
          + 0.5 (1-  h^z_{\tau^+}) (1+ h^z_{\tau^-}) =wt_{spin}^+ + wt_{spin}^- \label{eq:wtspinner3}
\end{equation}
For decay of scalar into $\tau\tau$ pair, the spin state probability $p_{\tau}^z=\frac{1}{2}$ (the subscript $z$ denotes spin projection on $z$ axes of  
frames type $C$). For ultrarelativistic cases, the $p_{\tau}^z$ means also probability of $\tau$ helicity states. 
For the decay of vector resonance, the  probability  $p_{\tau}^z$ depends on intermediate boson virtuality and $\tau$ direction,
we can  
rewrite formula  (\ref{eq:wtspinner2}) as
\begin{equation}
wt_{spin} =   p_{\tau}^z (1+  h^z_{\tau^+}) (1+h^z_{\tau^-})
          + (1- p_{\tau}^z) (1-  h^z_{\tau^+}) (1- h^z_{\tau^-}) =  wt_{spin}^+ + wt_{spin}^- \label{eq:wtspinner4}
\end{equation}

Note that $p_{\tau}^z$ can be interpreted as a
probability of creating {\ZW first} $\tau$ of the pair in a spin state $(+)$ along $z$ direction, whereas $P_{\tau}^z$
as its longitudinal polarization. 

Obviously, from comparison of above eqs. (\ref{eq:wtspinner2}) and (\ref{eq:wtspinner4}), following relation occur:  
\begin{equation}
p^z_{\tau}  =0.5 - 0.5 P_{\tau}^z  \label{eq:probl}
\end{equation}
which can be rewritten as
\begin{equation}
P^z_{\tau}  = 2 (1- p^z_{\tau} ) - 1.\label{eq:pol}
\end{equation}

 
\subsection{Longitudinal versus transverse spin correlations}

In the formalism discussed above, there is seemingly rather minor difference between situation when complete spin effects are
taken into account or only longitudinal ones. The case of
longitudinal spin effects only, means that  all $R_{i,j}$  components are set to zero except  
 $R_{t,z}$, $R_{z,t}$ and $R_{t,t}=1$, $R_{z,z} = \pm 1$; the sign depends whether decaying object is a scalar or a vector
and $R_{t,z}$ is equal to longitudinal $\tau$ polarisation $P_{\tau}^z$. No difference in the re-weighting algorithm is needed
in our solution. The transverse spin correlations are introduced with non-zero
$R_{x,x}$, $R_{y,y}$ and also off-diagonal $R_{x,y}$, $R_{y,x}$, without any
 modifications of the $t$ and $z$ components.

\section{The $2 \to 2$ parton level processes} \label{sec:TwoToTwo} 

The complete event weight $WT$, is factorised into multiplicative components: production ($wt_{prod}$), decay 
($wt_{decay}^{\tau^\pm}$) and spin correlation/polarization ($wt_{spin}$): 
\begin{equation}
WT = wt_{prod} wt_{decay}^{\tau^+}wt_{decay}^{\tau^-} wt_{spin}. 
\end{equation}
If the sample is polarized the last  factor should read $wt_{spin\; new}/wt_{spin\; old}$. 

Implemented methods  calculate internally complete set of weights but
as a default one, return the
spin correlation/polarization weight, as most often used in the applications\footnote{ Calculation
  is invoked with the call to {\tt calculateWeightFromParticlesH} method. 
The code organization is a consequence of implemented tests, availability of tested methods and 
available benchmarks. Not all features were expected to be included at the beginning:

At first, only longitudinal spin effects were expected to be taken into account and 
for  $ 2 \to 2$ at parton level  processes only. That is why, methods from {\tt KORALZ} \cite{koralz4:1994} could have been 
used. Later processes  $ 2 \to 4$ with two jets in final state made original methods 
focused around calculation of $\tau$ polarization i.e. $R_{t,z}$ less
intuitive/convenient. 
Introduction of options featuring complete spin effects made such organization even 
less suitable. Still we preserve backward compatibility,  frame for
functioning code and stability of results
for accumulated over years tests and examples for use. }. Other weights are
returned by supplementary methods\footnote{In sum for the formula~(\ref{eq:prod}),
  for  configurations of the incoming partons of  flipped flavours
  is enough to change the  sign of the
  $\cos\theta$ used  e.g.  in formula (\ref{eq:sigb}) of Appendix~\ref{app:sigborn}.
}. 

\subsection{Production weight}

The  production weight $wt_{prod}$ which  is a sum over all configurations of 
$\tau$ leptons spin and flavours of incoming partons of  matrix element squared 
multiplied by PDF's is calculated in frame $A'$. 
The parton level angular  kinematic configuration  are usually
taken in $A$ frame. The $R_{i,j}$ are calculated in $A$ or $A'$ frame,
except the case when transverse spin effects are taken into account
and frame $B$ has to be used.



\subsection{Decay weights}

Calculation of the decay weights  $wt_{decay}^{\tau^+}$ and $wt_{decay}^{\tau^-}$
is a by-product  of polarimetric vectors of $h^i_{\tau^+}$  and  $h^j_{\tau^-}$ calculation.
This calculation is performed in frames $D_{\tau^\pm}$.
The methods of {\tt Tauola} used for  $h^j_{\tau^-}$
calculation, at the same time calculate decay matrix elements squared summed over the $\tau$ spin
which are used in $wt_{decay}^{\tau^\pm}$ .



\subsection{ Spin correlation matrix $R_{i,j}$}\label{sec:Rij}

In most cases the spin correlation matrix $R_{i,j}$ is calculated in frame $A$
or $A'$.
The frame $B$ is used 
if transverse spin effects are taken into account. 

Calculation of production weight and of spin correlation matrix $R_{i,j}$ are related with each other. 
It is important to realize that in the collinear approximation
\begin{equation}
R_{t,z}=\frac{\sigma^+_{prod} - \sigma^-_{prod}}{\sigma^+_{prod} + \sigma^-_{prod}} \label{eq:RfromME}
\end{equation}

where $\sigma^\pm_{prod}$  denotes the sum of matrix element squared multiplied by PDF's,
the contribution, where first $\tau$ has spin $\pm$. In this approximation
and for Eq.~(\ref{eq:RfromME}) difference between frames $A$ and $A'$ can be overcomed.

Note, that the weight  $wt_{prod}$ is obtaines as
 a by-product of  $R_{i,j}$ calculation,
in fact of its $R_{t,z}$, $R_{z,t}$ components\footnote{
Relation between $R_{t,z}$, $R_{z,t}$ and modules of amplitudes is transparent in case of calculation as of Section~\ref{sec:VBF}.
}. 
In case of using effective Born cross-section 
or directly formula using  vector/axial couplings and $Z/\gamma^*$ propagators (implemented in {\tt TauSpinner})
this connection is less transparent.

\subsubsection{Case of spin = 0 resonance }

The  $R_{z,z} = -1$ if  $\tau$ mass is small in comparison to $\tau$-pair virtuality. 
The $R_{x,x} = R_{y,y} = 1$  for the scalar Higgs boson and  $R_{x,x} = R_{y,y} = -1$ for the pseudoscalar, 
while the mixed scalar-pseudoscalar represents only a slightly more complicated case;
the formula includes scalar-pseudoscalar
mixing angle and non-zero off-diagonal  $R_{x,y},  R_{y,x}$ terms. This extension is important for simulation of Higgs
CP parity signatures. 

\subsubsection{Case of spin = 1 resonance (Drell-Yan)}

The  $R_{z,z} = 1$ for neutral vector boson, if  $\tau$ mass is small in comparison of 
$\tau$-pair virtuality. The non-zero off-diagonal terms are
\begin{equation}
R_{t,z} = R_{z,t} = 2 \cdot p_{\tau}^z-1
\end{equation}

At present implementation,
if electroweak  corrections are neglected, the transverse spin correlations for neutral vector
boson/photon exchange are neglected as well and   $R_{x,x} = R_{y,y} =  R_{x,y} =  R_{y,x} = 0$. We are simply missing the code for the Born level expressions
taking into account transverse spin effects. Fortunately,
if  EW corrections calculation is activated, thanks to ported from {\tt Tauola} tables of {\tt SANC} library \cite{Andonov:2008ga}
results, not only first order EW corrections become available,
but  transverse spin effects  as well.

\subsubsection{Case of spin = 2 resonance X (non SM Higgs)}
{\MB
In this case the  $R_{z,z} = 1$ as in  Drell-Yan case.
The non-zero off-diagonal terms are also
\begin{equation}
R_{t,z} = R_{z,t} = 2 \cdot p_{\tau}^z-1,
\end{equation}
the difference is that  $ p_{\tau}^z$ feature different dependence on $\tau$ lepton directions
defined in $\tau$ lepton pair rest frame $B$. It is described by spherical 
polynomial of the fourth order, and not of the second order as for Drell-Yan.
This case  was introduced with the paper \cite{Banerjee:2012ez} and
corresponding example of {\tt TauSpinner} application.
}

\subsection{Calculating $wt_{spin}$ weight}

For the $wt_{spin}$ calculation formulae (\ref{eq:wtspinner0}, \ref{eq:wtspinner1}) are available, however
unless required by the user, the default version
is to omit transverse spin correlations, hence put $ R_{x,x} = R_{y,y} = R_{x,y} = R_{y,x} = 0$,
effectively reducing calculations to formula (\ref{eq:wtspinner2}). 
In the following, we will exploit prepared in the previous sub-section
components of $R_{i,j}$ for the individual cases.

\subsubsection{Case of spin = 0 resonance (scalar Higgs)}

\begin{equation}
 wt_{spin} =   1.0 +  sign \cdot h^{z}_{-}  h^{z}_{-} 
           +  R_{x,x} h^{x}_{+} h^{x}_{-} +  R_{yy} h^{y}_{+} h^{y}_{-} +  R_{x,y} h^{x}_{+} h^{y}_{-} +  R_{y,x} h^{y}_{+} h^{x}_{-}
\end{equation}

where $sign = -1$.

\subsubsection{Case of spin = 2 resonance X (non SM Higgs)}

\begin{equation}
wt_{spin} =  1.0 +  sign \cdot h^{z}_{+} h^{z}_{-} +  h^{z}_{+} +  h^{z}_{-}  + P^z_{\tau} h^{z}_{+} +  P^z_{\tau} h^{z}_{-} 
              +  R_{x,x} h^{x}_{+} h^{x}_{-} +  R_{y,y} h^{y}_{+} h^{y}_{-} +  R_{x,y} h^{x}_{+} h^{y}_{-} +  R_{y,x} h^{y}_{+} h^{x}_{-}
\end{equation}

where $sign = +1$. Note, that this form depends on  Born level distribution provided at the execution time
with  the  user function, {\MB which is used in calculating $R_{i,j}$,
at present in our examples  only $P_\tau^z$ is non-zero.}

\subsubsection{Case of spin = 1 resonance (Drell-Yan)} \label{sec:Spincases}

\begin{equation}
wt_{spin} = 1.0   +  sign \cdot h^{z}_{+} h^{z}_{-} +  h^{z}_{+} +  h^{z}_{-}  + P^z_{\tau} h^{z}_{+} +  P^z_{\tau} h^{z}_{-} 
                +  R_{1,1}^{EW} R_{x,x} h^{x}_{+} h^{x}_{-} + R_{2,2}^{EW} R_{y,y} h^{y}_{+} h^{y}_{-} 
                +  R_{1,2}^{EW} R_{x,y} h^{x}_{+} h^{y}_{-} + R_{2,1}^{EW} R_{y,x} h^{y}_{+} h^{x}_{-} \label{wt-spin}
\end{equation}

where $sign = +1$ and $R_{1,1}^{EW}$, $R_{1,2}^{EW}$, $R_{2,1}^{EW}$ and  $R_{2,2}^{EW}$ denote averaged (with production cross-section)
pre-tabulated transverse spin correlations terms, calculated using $\cal O(\alpha)$ EW corrections from SANC 
program\footnote{Nothing of principle prevents us to perform similar arrangements for $R_{t,z}$, that is for  $P^z_{\tau}$ which can be taken from 
  pre-tabulated electroweak results as well. Discussion of the systematic error
  is then nonetheless needed.}.
The $ R_{x,x} = R_{y,y} = R_{x,y} = R_{y,x} = 1$ should be set, in case we want to use 
$\cal O(\alpha)$ EW corrections from SANC. Or one can also set some of them  
to 1 and remaining ones to 0
in case one is  interested in studying only some components of  
transverse-spin effects.

The $ R_{x,x}, R_{y,y}, R_{x,y}, R_{y,x} $ are used directly as transverse
components
 for $R_{i,j}$ in case of non-standard model calculation. 
If one is   not interested in the actual size
of transverse effects, as predicted 
by $\cal O(\alpha)$ EW corrections, one can also modify hard coded pre initialized to zero values, or use discussed previously non SM Higgs options.
%

Generated Drell-Yan samples may feature only partial spin effects, e.g.
just  $ (1.0   +  sign \cdot h^{z}_{+} h^{z}_{-})$ part  of the spin effects.
Then  $wt_{spin}$  has to take this
into account, the denominator of such form has to be introduced,
to
 remove this partial effect. The numerator, given in eq. (\ref{wt-spin}),
will introduce the required (full) spin effect. Let us list the formulae
for the typical cases:

\begin{itemize}
\item 
If in the generated sample  spin correlations between two $\tau$ leptons are included but no other effects, the weight should take the form:
\begin{equation}
  wt_{spin}^{no\ pol} = \frac{wt_{spin}}{ 1.0   +  sign \cdot h^{z}_{+} h^{z}_{-} }.
  \label{eq:nopol}
\end{equation}
Such choice for the sample introduces dominant spin effect (the correlation),
which at the same time is free of systematic errors. It is a consequence
of the spin 1  state decaying to $\tau$ lepton pair. 
\item 
If in the generated sample spin correlations and  partial (no angular dependence and no incoming quark flavour dependence) 
polarization effects are included,  the weight should take the form:
\begin{equation}
wt_{spin}^{no\ angular} = \frac{wt_{spin}}{ 1.0   +  sign \cdot h^{z}_{+} h^{z}_{-} + P^{z, part}_{\tau} h^{z}_{+} +  P^{z, part}_{\tau} h^{z}_{-}} \label{eq:noang}
\end{equation}
where the denominator again represents spin correlations present in the sample.
One can expect
the  $P^{z, part}_{\tau}=P^{z}_{\tau}(s,\cos (\theta=0))$ calculated for incoming 
electrons (incoming electron, because it is the easiest way to
obtain average correction) reproduces correctly average individual $\tau$ 
polarization for events of $\tau$ pair virtuality close to the $Z$ boson peak. 
Such a choice for the generated sample could be motivated as follows. 
It features all spin effects except those which 
 depend on incoming partons PDF's, thus 
is free of related systematic errors, making it  convenient for studying systematic errors from PDFs. The corrections introduced with $wt_{spin}^{no\ angular}$ are even smaller than those of  $wt_{spin}^{no\ pol}$.
\end{itemize}

\subsection{Use of spin correlation/polarisation weight in special cases}

If generated sample feature spin effects then the   $wt_{spin}$ of formula (\ref{eq:useful}) can be used  as
weight $ WT = 1/wt_{spin}$ to remove spin effects from the generated sample. 

The cases when only part  of spin effects is taken into account, like in 
~\cite{Pierzchala:2001gc},  more specifically the
spin correlation  but no effects due to  vector and axial couplings to 
the intermediate $Z/\gamma^{*}$ state,
can be corrected with the help of the appropriate  weights as in examples for
Eqs. (\ref{eq:nopol}, \ref{eq:noang}). On the other hand, such partial
spin effects can be also removed completely, e.g. for consistency checks.
Also the case, when spin correlations and average over the $\tau$
lepton angles within rest-frame of lepton pair are present, but no angular dependence is taken into account, can be checked in an analogous way.

\subsection{Born cross-section}

Until now, we were expressing formulae in the language of spin amplitudes and spin weights (spin correlation matrix
$R_{i,j}$). However, often cross-sections calculated at the parton level are available for the explicit helicity
states of the outgoing $\tau$ leptons.  
Several such lowest order formulae for calculating cross-sections of an analytical form, are  implemented in the code  
of the {\tt TauSpinner} itself, or of its use examples. Some of them represent {\it legacy} code, ported from 
different projects, others were
coded for {\tt TauSpinner} application as benchmarks on the EW parameters 
setting and are used in the published tests. For completness of documentation
we  decribe them in Appendix \ref{app:sigborn}.

\section{Polarimetric vectors, polarization and helicity states} \label{sec:HHcalc}

Calculation of the polarimetric vectors $h_{\tau}$ is performed using code from {\tt Tauola} library. 
The conventions is that  $h_{\tau} = (h_{\tau}^x, h_{\tau}^y, h_{\tau}^z, h_{\tau}^t)$ (i.e. last component is time-like) following 
convention used in {\tt FORTRAN}. The same {\tt FORTRAN} code calculates decay matrix element squared, 
which are returned as well, as auxiliary information.

Depending on which decay channel is present, calculation is performed using respective functions
from {\tt Tauola} library.
Not for all $\tau$ decay channels, see Table~\ref{Tab:BranchingFrac},
 $h_\tau$ is  nonetheless calculated for 97.5\% of the decay width is
covered, i.e. polarimetric vector $h$ is calculated, otherwise its spatial
components are set to 0. This is not the case for {\tt Tauola} library,
if used for generating $\tau$ decays. Then only for channels featuring more
than 5 pions polarimetric vector is calculated with approximation. Still then,
it is not set to zero.

To calculate decay matrix elements, $\tau$ decay products have to be boosted into $\tau$ rest-frame: we first boost all products
to $\tau-\tau$ pair rest frame, then rotate from $A$ to $B$ so that $\tau$ of $Z/H$ decay is along $z$-axis, intermediate step is boost from the  $\tau-\tau$ pair
rest-frame to  $C_\tau$, 
then rotation to have neutrino from $\tau$ decay along $z$ axis ($D_\tau$ frame);
calculated for that purpose  angles $\phi_2$, $\theta_2$ 
are stored for rotating back of the polarimetric vector to the $\tau$
rest-frame $B$ or $C_\tau$.
\begin{table}
\begin{center}                               
\begin{tabular}{|l|r|}
\hline \hline
$\tau$ decay mode          & Branching fractions \% \\
\hline 
$e^- \bar{\nu}_e \nu_{\tau}$                            &       17.85          \\
$\mu^- \bar{\nu}_{\mu} \nu_{\tau}$                       &       17.36          \\
$\pi^- \nu$                                           &       10.91          \\
$\pi^- \pi^{0} \nu$                                    &       25.51          \\
$\pi^- \pi^{0} \pi^{0} \nu, \pi^- \pi^+ \pi^- \nu $     &        9.29, 9.03 (incl. $\omega$)          \\
$ K^- \nu$                                            &       0.70          \\
$ K^- \pi^{0} \nu,  K^0 \pi^- \nu$                     &       0.43, 0.84          \\
$\pi^- \pi^+ \pi^- \pi^0  \nu$                         &       4.48 (incl. $\omega$)          \\
$\pi^- \pi^0 \pi^0 \pi^0  \nu$                         &       1.04          \\
Other                                                 &       2.5   \\
\hline 
\end{tabular}
\end{center}
\caption{ Summary of $\tau$ decay modes implemented for calculation of polarimetric vectors, 
  branching fraction for each mode accordingly to~\cite{Nakamura:2010zzi}
  illustrate completnes of the algorithm. The ``Other'' decay modes  are treated 
as unpolarized ones.
 \label{Tab:BranchingFrac} }
\end{table}

As a by-product of $h_\tau$  calculations, 
 $wt_{decay}^{\tau^\pm}$ of Eq.~(\ref{eq:dec}) is obtained.



\subsection{Longitudinal polarization}

Calculation of components of $R_{i,j}$ spin density matrix of $\tau$-pair
production in ultrarelativistic limit simplifies, it
is then equivalet  to calculation of modules for matrix elements of 
$\tau$-pair production in a given helicity configurations.
Depending on the approach, either $p^z_{\tau}$ or $P^z_{\tau}$ is calculated from the other one, using respectively formula
(\ref{eq:pol}) or (\ref{eq:probl}).
The case, when electroweak effects are included, technically differs only slightly%
\footnote{
\ZW The same formulae using $R_{i,j}$ matrix are still used, individual parton flavour
contributions to $R_{i,j}$ are 
obtained numerically from the   
 pre-tabulated values  
for some  invariant masses and scattering angles. These pre-tabulated values 
are read from the stored files.  Then results are { interpolated} to obtain results for required
arguments.  The $x$ and $y$ components of  $R_{i,j}$ are pre-tabulated and available for use.},
despite the fact that then the  transverse spin effects can be taken into account.
In all cases the relation $R_{t,z}= P^z_{\tau}=$ holds, {\ZW  no pretabulated electroweak
  results are  used for this component of  $R_{i,j}$. This can be easily activated though, e.g. if
  results for large $s$ are needed. Around the $Z$ peak the presently implemented electroweak
library may be less suitable.}
 Formula  (\ref{eq:wtspinner4}) is used for attribution of helicity states. 

\subsubsection{Case of spin = 0 resonance (scalar (Higgs)}
In $H \to \tau \tau$ decays, the probability of the helicity state  denoted $p^z_{\tau}$ is equal to 0.5 for $ P^z_{\tau} = 1$ and for $ P^z_{\tau} = -1$ as well, 
see Table~\ref{Tab:ProbHelicity}.

\subsubsection{Case of spin = 1 resonance (Drell-Yan)}

In $Z/\gamma^* \to \tau \tau$ decays, the probability of the helicity state  denoted $p^z_{\tau}$  is a function of the $\tau$ scattering angle, 
$\theta$, and the center of mass squared of the hard process, $s$. It is true at the Born level, and in
ultrarelativistic limit \cite{Pierzchala:2001gc}.

Note that
the rest frame of the production process and the rest frame of the $\tau$-pair
might not be the same due to photon 
bremsstrahlung in the $\tau$ pair production vertex. We have discussed this already in Section \ref{sec:frames}, but let us
complete some more technical details now.

The center-of-mass system of $\tau$-pairs is used for calculating $ \cos \theta$ angle 
and the direction of incoming partons is set along $z$-direction in the laboratory frame.
Following definitions are used:
\begin{equation}
\cos \theta_{1} = \frac{\vec{\tau}^+ \cdot \vec{b}^{+}}{|\vec{\tau}^{+}| |\vec{b}^{+}|},    \ \ \ 
\cos \theta_{2} = \frac{\vec{\tau}^- \cdot \vec{b}^{-}}{|\vec{\tau}^{-}| |\vec{b}^{-}|},   
\end{equation}
where $ \vec{b}^{+} $, $ \vec{b}^{-}$ denote 3-vectors of incoming beams, in the laboratory frame defined along the z-axis.
And the definition of {\it effective} $\cos \theta$ is constructed following Ref.~\cite{Was:1989ce}
\begin{equation}
\cos \theta = \frac{\cos \theta_{1} \cdot \sin \theta_{2} + \cos \theta_{2} \cdot \sin \theta_{1} }{ \sin \theta_{1} +  \sin \theta_{2}}.
\end{equation}

At the Born level, for a given flavour of incoming partons and in the ultrarelativistic limit, 
the $p^z_{\tau}$ is defined \cite{Pierzchala:2001gc} as

\begin{equation}
p^z_{\tau}  =\frac { \frac {d \sigma}{d \cos \theta}(s, \cos \theta, P^z_{\tau} = 1)}
          { \frac {d \sigma}{d \cos \theta}(s, \cos \theta, P^z_{\tau} = 1) + \frac {d \sigma} {d \cos \theta}(s, \cos \theta, P^z_{\tau} = -1) }, 
\end{equation}

where

\begin{equation}
\frac{d \sigma}{d \cos \theta}(s, \cos \theta, P^z_{\tau}) =  (1 + \cos^2 \theta)F_0(s) + 2 \cos \theta\ F_1(s) 
                                              -  P^z_{\tau} [(1 + \cos^2 \theta)F_2(s) + 2 \cos \theta\ F_3(s)] 
\end{equation}

and $F_i(s)$ are four form factors which depend on the initial and the final state fermion couplings to the Z boson and the 
propagator~\cite{Pierzchala:2001gc}. 

In case of {\tt TauSpinner} algorithm flavour of the initial partons is
assumed to be unknown, the probability  $p^z_{\tau}$  is calculated as a weighted average over all possible
initial state quark configurations. This requires averaging (summing)  over all production matrix elements square convoluted with 
the structure functions
\begin{equation}
   p^z_{\tau} = \frac{\sum_{(flavours: i,j)} f_i(x_1,Q^2) f_j(x_2, Q^2) \cdot \sum_{spin }|{\cal M}^{prod}_{i,j}|^2 \cdot p^z_{\tau}(i,j)}
                   {\sum_{(flavours: i,j)} f_i(x_1,Q^2) f_j(x_2, Q^2) \cdot\sum_{spin }|{\cal M}^{prod}_{i,j}|^2 },
\end{equation}
where indices i,j denoting flavours of incoming partons are now explicitly written.

For calculating the spin weight, the polarisation $P^z_{\tau}$ of the single $\tau$ in a mixed quantum state, 
is calculated neglecting transverse spin degrees of freedom (ultrarelativistic limit).
The $P^z_{\tau}$ is then linear
function of probability for the helicity state\footnote{ 
Note that in paper \cite{Czyczula:2012ny} formula  $ P_{\tau}^z=(2 p^z_\tau-1)$ is used instead of  Eq.(\ref{eq:pol}). 
There, different (rotated by $\pi$ angle with respect to axis perpendicular
to reaction frame is used. Similar convention  matching 
may be necessary in programs as well, see e.g. {\tt  nonSM\_adopt(); nonSM\_adoptH()} methods of {\tt tau-reweight-test.cxx}. 
}, Eq.(\ref{eq:pol}).


\subsubsection{Case of spin = 2 resonance X (not a Higgs)}
{\ZW
  In this case user provided function is required to replace $\frac {d \sigma}{d \cos \theta}$,
  the Non-Standard physics 
effective Born parametrizations are not included in the {\tt TauSpinner} 
library, dummy function is provided only. It can be replaced at 
the execution time with the user one,  see Ref.~\cite{Banerjee:2012ez} 
for details 
and in particular
{\tt TAUOLA/TauSpinner/examples/ tau-reweight-test.cxx } directory. 
This $2\to2$ parton level process
function {\tt nonSM\_adoptH} (or {\tt nonSM\_adopt}) has the 
following arguments  
{\tt  (int ID, double S, double cost, int H1, int H2, int key)}.
It can be activated  in the user code at the execution time, see Appendix~\ref{app:ME}.
}

\subsection{Attributing $\tau$ (or $\tau$ pair) helicity states}

For the individual event, the  $\tau$ helicity ($+/-$) is attributed 
stochastically. The probability for given configurations of the longitudinal
polarisation of $\tau$'s of different origins~\cite{Pierzchala:2001gc} is shown in Table~\ref{Tab:ProbHelicity}.
These probabilities can be used, when  decay kinematical configurations of the
decays are not yet constructed.

\begin{table}
\vspace{2mm}
\begin{center}                               
\begin{tabular}{|l|r|r|r|}
\hline \hline
Origin                               & Helicity ${\tau_1}$ & Helicity  ${\tau_2}$ & Probability \\
\hline 
Neutral Higgs boson: $ H $            & +   & -   & 0.5 \\
                                      & -   & +   & 0.5 \\
Neutral vector boson: $Z/\gamma^*$    & +   & +   & $p_{\tau}^{z}$ \\
                                      & -   & -   & 1 - $p_{\tau}^{z}$ \\
Charged Higgs boson:  $H^{\pm}$        & +   & -   & 1.0 \\
Charged vector boson: $W^{\pm}$        & -   & -   & 1.0 \\
\hline 
\end{tabular}
\end{center}
\caption{ Probability for the helicities  of the  of $\tau$'s from different origins~\cite{Pierzchala:2001gc}. 
 \label{Tab:ProbHelicity} }
\end{table}

However, once the  configurations of decay products are generated,
event-by-event calculation of the probability for helicity states,
has to take into account $\tau$ decay's
polarimetric vectors $h_{+}$ and  $h_{-}$.
 The  $(+)$ is chosen if 
probabilities, calculated with the formulae below, are smaller than generated random
number  {\tt RRR}, otherwise $(-)$ is taken: 

\begin{itemize}
\item{\bf Case of spin = 0 resonance (scalar Higgs)}
\begin{equation}
 p(-) = \frac { 1 +  sign \cdot h^{z}_{+} \cdot h^{z}_{-} +  h^{z}_{+} +  h^{z}_{-} }
              { 2 +  2 \cdot sign \cdot h^{z}_{+} \cdot h^{z}_{-}  }
\end{equation}
where $sign = -1$.
\item{\bf Case of spin = 2 resonance X (not a Higgs)}
\begin{equation}
p(-) = (1 + P^z_{\tau}) \frac { 1 +  sign \cdot h^{z}_{+} \cdot h^{z}_{-} +  h^{z}_{+} +  h^{z}_{-} }
                             { 2 +  2 \cdot sign \cdot h^{z}_{+} \cdot h^{z}_{-} + 2.0 \cdot P^z_{\tau} \cdot (h^{z}_{+} +  h^{z}_{-}) }
\end{equation}
where $sign = +1$.
\item{\bf Case of spin = 1 resonance (Drell-Yan)}
\begin{equation}
p(-) = (1 + P^z_{\tau}) \frac { 1 +  sign \cdot h^{z}_{+} \cdot h^{z}_{-} +  h^{z}_{+} +  h^{z}_{-} }
                             { 2 +  2 \cdot sign \cdot h^{z}_{+} \cdot h^{z}_{-} + 2.0 \cdot P^z_{\tau} \cdot (h^{z}_{+} +  h^{z}_{-}) }
\end{equation}
where $sign = +1$.
\end{itemize}

\section{The $2 \to 4$  process} \label{sec:VBF}

At this point we have completed discussion
for the case when 
{\tt TauSpinner} is identifying and working with parton level processes 
of $2\to 2$ type. Let us turn now to the case when $2\to 4$ parton level processes
are used. This extension has been introduced very recently, see \cite{Kalinowski:2016qcd} for detailed
description. 


A baseline algorithm remains essentially the same.  The differences can be grouped into few parts
\begin{itemize}
\item {\bf Kinematics} \\
 For calculation of factorisation scale, instead of virtuality 
of $\tau$ lepton pair, virtuality of lepton pair combined with jets is used. 
In fact few options to define the $Q^2$ factorisation scale have been implemented. 
The  definitions 
of frames type $A$, $B$, $C_{\tau^\pm}$ and  $D_{\tau^\pm}$  as explained for 
case of $2 \to 2$ 
processes are used. In principle an additional rotation by 
an angle $\pi$ around axis $x$ would be again needed  for the $\tau^-$
in transformation
from $B$, $C_{\tau^-}$ frames. This is  due to  convention used for
spin amplitudes\footnote{ In practice, as no transverse spin effects are taken into account, the convention change
  is performed while  calculating
 {\tt double W[2][2] } (amplitudes squared for the helicity states) in {\tt wbfdistr.cxx}. }
Definition of  the previously discussed frame $A'$  needs to be modified.
Not only bremsstrahlung photons  
need to be integrated into  $\tau$ leptons to obtain kinematic configurations
for matrix element calculation, but outgoing partons (jets) four-momenta have to be taken into
account as well.

\item{\bf Matrix Element Calculation} \\ Matrix element squared are calculated
for given helicity states of $\tau$ leptons and flavours of 
incoming/outgoing partons in frame A'. In particular formula
(\ref{eq:RfromME}) is used for calculation of $R_{z,t}$ and $R_{t,z}$, which are
then used as of frame B,  without any modifications.
The EW scheme used and $\alpha_s(Q^2)$ can be  configured using several
options defined in Ref.~\cite{Kalinowski:2016qcd}. 
\item{\bf Extension due to flavours of accompanying partons/jets}\\
All sums $\sum_{flavour}$ as defined in Section~\ref{sec:partonlevel} should
read now as sums over flavours of incoming {\it and} outgoing partons.
Of course PDFs distributions are used for incoming partons only.
{\ZW This modification
is  extending re-weighting algoritm dependence on final state components
consisting of partons (hadronic jets). This technically simple change may be thus of a concern
because of additional types of the systematic errors involved.
}
\end{itemize}

Except these changes, algorithms remain not modified.
However only longitudinal spin effects can be simulated because
we use helicity states, that is  ultrarelativistic 
approximation in matrix element calculation. In future,
if necessary, we can extend the case to complete 
spin effects treatment. Only minor extension of the program will be needed.
Approximation is in the matrix element calculation only. It may even be
possible to use for that purpose presently available  {\tt MadGraph} amplitudes
discussed in~\cite{Kalinowski:2016qcd}.
Better control of the complex phases will be however needed.

\section{Summary} \label{sec:Summary}


In this paper we have presented theoretical basis and description of algorithms used in the {\tt TauSpinner} program for simulating
spin effects in the production and decay of $\tau$ leptons in proton-proton collisions at LHC. 
The primary source of $\tau$ leptons are Drell-Yan processes of single $W$ and $Z$ boson production, 
often accompanied by the additional jets. Since Higgs boson discovery by ATLAS and CMS Collaboration in 2012,
the $H \to \tau \tau$ decay became the most promising channel for studying CP properties of its couplings 
to fermions. With more data available from Run II of LHC also $\tau$ decays in case of multi-boson production will become
part of  interesting signatures. 

Discussed here algorithms allow for separation of the $\tau$ lepton production and decay
in phenomenological studies. We have discussed 
options used for simplifying descriptions and approximations used.

Although the framework of {\tt TauSpinner} is prepared for implementation of the
non Standard Model couplings, we have not 
given too much attention to this possible development path. 
%
%
In the paper appendices, instead, we have  recalled all available in the distribution  examples and applications
and pointed to  documenting them publications, wherever possible. The technical
information given here is very brief, detailes are delegated
to the {\tt README} files available with the distributed code.

\vskip 1 cm
\centerline{\bf \Large Acknowledgments}
\vskip 0.5 cm
We would like to acknowledge and thank several colleagues who either contributed directly or 
by stimulating discussions to the devellopmement of the {\tt TauSpinner} package and its applications: M. Bahmani,
S. Banerjee, Z. Czyczula, W. Davey, A. Kaczmarska, J. Kalinowski and W. Kotlarski. 

\begingroup\begin{thebibliography}{10}

\bibitem{Czyczula:2012ny}
Z.~Czyczula, T.~Przedzinski, and Z.~Was, {\em Eur.Phys.J.} {\bf C72} (2012)
  1988,
\href{http://www.arXiv.org/abs/1201.0117}{{\tt 1201.0117}}.

\bibitem{Banerjee:2012ez}
S.~Banerjee, J.~Kalinowski, W.~Kotlarski, T.~Przedzinski, and Z.~Was, {\em
  Eur.Phys.J.} {\bf C73} (2013) 2313,
\href{http://www.arXiv.org/abs/1212.2873}{{\tt 1212.2873}}.

\bibitem{Kaczmarska:2014eoa}
A.~Kaczmarska, J.~Piatlicki, T.~Przedzinski, E.~Richter-Was, and Z.~Was, {\em
  Acta Phys. Polon.} {\bf B45} (2014), no.~10 1921--1946,
\href{http://www.arXiv.org/abs/1402.2068}{{\tt 1402.2068}}.

\bibitem{Przedzinski:2014pla}
T.~Przedzinski, E.~Richter-Was, and Z.~Was, {\em Eur.Phys.J.} {\bf C74} (2014)
  3177,
\href{http://www.arXiv.org/abs/1406.1647}{{\tt 1406.1647}}.

\bibitem{Kalinowski:2016qcd}
J.~Kalinowski, W.~Kotlarski, E.~Richter-Was, and Z.~Was, {\em Eur. Phys. J.}
  {\bf C76} (2016), no.~10 540,
\href{http://www.arXiv.org/abs/1604.00964}{{\tt 1604.00964}}.

\bibitem{Davidson:2010rw}
N.~Davidson, G.~Nanava, T.~Przedzinski, E.~Richter-Was, and Z.~Was, {\em
  Comput.Phys.Commun.} {\bf 183} (2012) 821--843,
\href{http://www.arXiv.org/abs/1002.0543}{{\tt 1002.0543}}.

\bibitem{Jadach:1999vf}
S.~Jadach, B.~Ward, and Z.~Was, {\em Comput.Phys.Commun.} {\bf 130} (2000)
  260--325,
\href{http://www.arXiv.org/abs/hep-ph/9912214}{{\tt hep-ph/9912214}}.

\bibitem{Jadach:1998wp}
S.~Jadach, B.~F.~L. Ward, and Z.~Was, {\em Eur. Phys. J.} {\bf C22} (2001)
  423--430,
\href{http://www.arXiv.org/abs/hep-ph/9905452}{{\tt hep-ph/9905452}}.

\bibitem{jadach-was:1984}
S.~Jadach and Z.~W\c{a}s, {\em Acta Phys. Polon.} {\bf B15} (1984) 1151,
  \uppercase{E}rratum: {\bf B16} (1985) 483.

\bibitem{Jadach:1993hs}
S.~Jadach, Z.~W\c{a}s, R.~Decker, and J.~H. K\"{uhn}, {\em Comput. Phys.
  Commun.} {\bf 76} (1993)
361.

\bibitem{Davidson:2010ew}
N.~Davidson, T.~Przedzinski, and Z.~Was, {\em Comput. Phys. Commun.} {\bf 199}
  (2016) 86--101,
\href{http://www.arXiv.org/abs/1011.0937}{{\tt 1011.0937}}.

\bibitem{Nanava:2009vg}
G.~Nanava, Q.~Xu, and Z.~Was, {\em Eur. Phys. J.} {\bf C70} (2010) 673--688,
\href{http://www.arXiv.org/abs/0906.4052}{{\tt 0906.4052}}.

\bibitem{Richter-Was:2016mal}
E.~Richter-Was and Z.~Was, {\em Eur. Phys. J.} {\bf C76} (2016), no.~8 473,
\href{http://www.arXiv.org/abs/1605.05450}{{\tt 1605.05450}}.

\bibitem{Richter-Was:2016avq}
E.~Richter-Was and Z.~Was, {\em Eur. Phys. J.} {\bf C77} (2017), no.~2 111,
\href{http://www.arXiv.org/abs/1609.02536}{{\tt 1609.02536}}.

\bibitem{koralz4:1994}
S.~Jadach, B.~F.~L. Ward, and Z.~W\c{a}s, {\em Comput. Phys. Commun.} {\bf 79}
  (1994) 503.

\bibitem{Andonov:2008ga}
A.~Andonov {\em et al.}, {\em Comput. Phys. Commun.} {\bf 181} (2010) 305--312,
\href{http://www.arXiv.org/abs/0812.4207}{{\tt 0812.4207}}.

\bibitem{Pierzchala:2001gc}
T.~Pierzchala, E.~Richter-Was, Z.~Was, and M.~Worek, {\em Acta Phys.Polon.}
  {\bf B32} (2001) 1277--1296,
\href{http://www.arXiv.org/abs/hep-ph/0101311}{{\tt hep-ph/0101311}}.

\bibitem{Nakamura:2010zzi}
{Particle Data Group} Collaboration, K.~Nakamura {\em et al.}, {\em J. Phys.}
  {\bf G37} (2010)
075021.

\bibitem{Was:1989ce}
Z.~Was and S.~Jadach, {\em Phys. Rev.} {\bf D41} (1990)
1425.

\bibitem{Alwall:2006yp}
J.~Alwall {\em et al.}, {\em Comput. Phys. Commun.} {\bf 176} (2007) 300--304,
\href{http://www.arXiv.org/abs/hep-ph/0609017}{{\tt hep-ph/0609017}}.

\bibitem{Dobbs:2001ck}
M.~Dobbs and J.~B. Hansen, {\em Comput. Phys. Commun.} {\bf 134} (2001) 41--46,
https://savannah.cern.ch/projects/hepmc/.

\bibitem{Golonka:2002rz}
P.~Golonka, T.~Pierzchala, and Z.~Was, {\em Comput. Phys. Commun.} {\bf 157}
  (2004) 39--62,
\href{http://www.arXiv.org/abs/hep-ph/0210252}{{\tt hep-ph/0210252}}.

\bibitem{Davidson:2008ma}
N.~Davidson, P.~Golonka, T.~Przedzinski, and Z.~Was, {\em Comput. Phys.
  Commun.} {\bf 182} (2011) 779--789,
\href{http://www.arXiv.org/abs/0812.3215}{{\tt 0812.3215}}.

\bibitem{Jozefowicz:2016kvz}
R.~Jozefowicz, E.~Richter-Was, and Z.~Was, {\em Phys. Rev.} {\bf D94} (2016),
  no.~9 093001,
\href{http://www.arXiv.org/abs/1608.02609}{{\tt 1608.02609}}.

\bibitem{Barberio:2017ngd}
E.~Barberio, B.~Le, E.~Richter-Was, Z.~Was, D.~Zanzi, and J.~Zaremba, {\em
  Phys. Rev.} {\bf D96} (2017), no.~7 073002,
\href{http://www.arXiv.org/abs/1706.07983}{{\tt 1706.07983}}.

\bibitem{Bahmani:2017wbm}
M.~Bahmani, J.~Kalinowski, W.~Kotlarski, E.~Richter-Was, and Z.~Was, {\em Eur.
  Phys. J.} {\bf C78} (2018), no.~1 10,
\href{http://www.arXiv.org/abs/1708.03671}{{\tt 1708.03671}}.

\bibitem{Andonov:2004hi}
A.~Andonov {\em et al.}, {\em Comput. Phys. Commun.} {\bf 174} (2006) 481--517,
\href{http://www.arXiv.org/abs/hep-ph/0411186}{{\tt hep-ph/0411186}}.

\end{thebibliography}\endgroup
\providecommand{\href}[2]{#2}
\appendix

\section{Project software organization}
The code of {\tt TauSpinner} is distributed as a part of  the 
{\tt Tauolapp} \cite{Davidson:2010rw} package.
The main modules of the library source code reside in {\tt /src} and {\tt /src/VBF} 
sub directories of
{\tt TAUOLA/TauSpinner}. The compiled libraries are placed in 
{\tt TAUOLA/TauSpinner/lib} and copied to {\tt TAUOLA/lib} directory
during installation of {\tt TAUOLA} library with {\tt TauSpinner}.
Technical details on installation and execution of main examples 
are provided in {\tt README} files of subdirectories: \\
{\tt TAUOLA/TauSpinner},  \\
{\tt TAUOLA/TauSpinner/examples}, \\
 {\tt TAUOLA/TauSpinner/examples/example-VBF}, \\
 {\tt TAUOLA/TauSpinner/examples/example-VBF/SPIN2}. \\
The {\tt changelog.txt} file is also provided, however, since November 2012
logs are set into the  {\tt TAUOLA changelog.txt}. Note that 
the file {\tt html/index.html} is of the old {\tt TauSpinner} versions 
prior to its merge with {\tt Tauolapp} distribution too.

The distribution includes extended set of examples and applications,
 collected in  {\tt TAUOLA/TauSpinner/examples } sub-directories:
\begin{itemize}
\item
{\tt CLEO-to-RCHL}, see App.~\ref{app:rchl},
\item
 {\tt CP-tests}, see Apps.~\ref{app:CP}~\ref{app:CP-h}~\ref{app:CP-z},
and also introduction to electroweak effects implementation, App.~\ref{app:electroweak}.
\item
{\tt example-VBF},  see App.~\ref{app:jj} and ~\ref{app:spin2},
\item
{\tt  testsERW }, see App.~\ref{app:other},
\item
 {\tt  applications},  general methods,
useful to deal with event record contents
and {\tt README} file is included only.
\item
 {\tt applications/applications-fits,} see App.~\ref{app:bench},
\item
{\tt applications/applications-plots-and-paper}, see App.~\ref{app:bench},
\item
{\tt applications/applications-rootfiles},  see App.~\ref{app:bench},
\item
{\tt applications/test-bornAFB},  see App.~\ref{app:bench},
\item
{\tt applications/test-ipol, } see App.~\ref{app:bench}.
\end{itemize}

Some of obtained with the program published results are stored in further
sub-directories of {\tt TAUOLA/TauSpinner}, that is in
\begin{itemize}
\item
{\tt    paper},  see App.~\ref{app:bench}, and Ref.~\cite{Czyczula:2012ny},
\item
 {\tt     paper-spin2},   see App.~\ref{app:ME}, and Ref.~\cite{Banerjee:2012ez},
\item
 {\tt     paper-spin2-higgs},   see App.~\ref{app:ME}, and Ref.~\cite{Banerjee:2012ez},
\item
 {\tt paper-application-studies},  see App.~\ref{app:bench}, and Ref.~\cite{Kaczmarska:2014eoa} \\ (redirects to  {\tt TAUOLA/TauSpinner/examples/applications/applications-plots-and-paper}),
\item
 {\tt paper-CP} see App.~\ref{app:CP} and Ref.~\cite{Przedzinski:2014pla}.   
\end{itemize}

There is no need for the user to read details of all this  multitude of 
 files, except the one of the interest.
\\
The code of the basic examples is stored in  the following two files: \\
{\tt TAUOLA/TauSpinner/examples/tau-reweight-test.cxx}  for $2\to 2$ processes,  \\
{\tt TAUOLA/TauSpinner/examples/example-VBF/example-VBF.cxx} for $2\to 4$ processes.

The methods needed to read events from datafiles are stored separately 
as they depend on particular  datafile format: \\ 
{\tt  TAUOLA/TauSpinner/examples/read\_particles\_from\_TAUOLA.cxx} for  $2\to 2$ processes,
\\ and 
{\tt TAUOLA/TauSpinner/examples/example-VBF/read\_particles\_for\_VBF.cxx} for the $2\to 4$ processes%
 \footnote{The {\tt lhef-to-hepmc.cxx}, include  program to convert
 necessary for {\tt TauSpinner} event content from {\tt LHEF}   format \cite{Alwall:2006yp} to 
{\tt HepMC} format \cite{Dobbs:2001ck}}.

The {\tt FORTRAN} examples  of the user functions for $2\to 2$ processes
{\tt  ggHXtautau.f}, {\tt distr.f} (accessible through the C++ wrapers)
are stored in {\tt TAUOLA/TauSpinner/examples/} and example
 library  for implementing user $2\to 4$ processes in
{\tt TAUOLA/TauSpinner/examples/example-VBF/SPIN2}.

\section{General information on user interface methods} \label{sec:methods}
Over the years several methods of initializing and operating {\tt TauSpinner} 
were developped~\cite{Czyczula:2012ny,Banerjee:2012ez,Kaczmarska:2014eoa,Przedzinski:2014pla,Kalinowski:2016qcd}.
Technical descriptions and discussions on systematic errors were provided, but remained scattered over several publications.
Collecting them or providing references from  one place is now of convenience.

The rather nontrivial component of analysis code is a correct reading and interpretation
of information in the event record,
namely finding outgoing $\tau$ leptons and their decay products\footnote{
  Also  bremsstrahlung photons from $\tau$-pair production process. In case of multitude entries for the single   particle,
  identyfication of  the one matching the kinematical constraints is necessary.  
  It often requires navigation back and forth through the event record, finding links to mothers, etc.
}. The standard of the {\tt HepMC}  format  \cite{Dobbs:2001ck} of the 
event record is a usual input format for {\tt TauSpinner},
but some helping classes are prepared in the examples, to allow also
for reading {\tt LHEF} format \cite{Alwall:2006yp}. These interface classes are usually kept separately from the examples code
and described 
in the appropriate {\tt README} files of the particular example directory. In several cases also  input
files with  small number of generated
events are provided together with the examples.

The technical implementation of the methods used in examples,
evolved during the last few
 years. Most of them 
 use  basic (sometimes also additional developped during the project)
 functionalites of the {\tt MCTester} package~\cite{Golonka:2002rz,Davidson:2008ma}
 and its native
 analysis function {\tt MC\_Analyze}. This allows to standarize not only production of benchmark histograms, but also
 validation of class reading event record,
 {\tt readParticlesFromTAUOLA\_HepMC} method. This method is expected to be adapted or
replaced by the user own, depending on the particular event record form.

Let us list some advantages of   {\tt MCTester}:
\begin{enumerate}
 \item {\tt MCTester} interface to  {\tt HepMC} is profoundly tested for
  many conventions on how to handle 
  {\tt HepMC} event record exceptions in the way it was filled in.
  {\tt MCTester} offers thus debugging facility for communication  too.
\item The only information exchanged between {\tt MCTester} and  {\tt TauSpinner} modules
 is 
the event weight which is passed to {\tt MC\_Analyze} function together with the  event itself \\
{\tt  MC\_Analyze(temp\_event,WT);}
\item Particular test specific plots, are defined in {\tt C} scripts with names
following {\tt MCTester} analysis template script
{\tt UserTreeAnalysis.C}. These user-defined plots are then generated by {\tt MC\_Analyze}
function.
\end{enumerate}

\noindent
 The {\tt TauSpinner} and
{\tt MCTester} codes are independent, use of the latter is not
mandatory, see Appendix~\ref{app:other}.

We are now ready to return to documentation of the  {\tt TauSpinner}
application methods.
There are three categories of methods prepared for the user: for initialization, evaluation of 
particular event and finally for gathering auxiliary information.
Let us start from general methods. Methods for the specific applications 
will be discussed later, case after case in sub-appendices.

\vskip 2mm
  \centerline{\bf Initialization}
\vskip 2mm

Before {\tt TauSpinner} is activated, and individual events analyzed, general
information about sample on which it is going to be used has to be provided.
Initialization of {\tt TauSpinner} has to be adjusted to the sample specific.
The initialization is divided into two categories, the general one
discussed now and the one specific
for particular application which will be discussed in the sub-appendices.
In every case the following initialization routine has to be invoked:

   {\tt \small  Tauola::initialize();} \\
to intialize $\tau$ decay library.
Generation of decays  will not be performed but constants necessary for
calculation of decay matrix elements have to be set.
\noindent
The library of PDFs
has to be initialized as well,

   {\tt \small  string name="MSTW2008nnlo90cl.LHgrid";  \\
   LHAPDF::initPDFSetByName(name);} \\
 for every application and finally the basic initialization method
of {\tt TauSpinner} can be invoked, 

   {\tt \small initialize\_spinner(Ipp, Ipol, nonSM2, nonSMN,  CMSENE);}\\
The input parameters of this method are as follows:
\begin{itemize}
\item 
  {\tt bool Ipp = true;} denote that program is to be used for $pp$ collisions.
  At present it is the only available option.
\item {\tt int Ipol; }  states of the polarization in the input sample
\begin{itemize}
\item  0 - events generated without spin effects
\item  1 - events generated with all spin effects
\item  2 - events generated with spin correlations only, no individual $\tau$
           polarization.
\item  3 - events generated with spin correlations and polarization but 
           missing angular dependence of individual $\tau$ polarization.
\end{itemize}
\item {\tt  int nonSM2 = 0} - Are we using nonSM calculations? y/n  (1/0),
\item {\tt int nonSMN = 0 } - If we are using nonSM calculations we may consider
                   corrections 
                   to shapes only: y/n  (1/0),
                   this option was introduced for $2 \to 2$ algorithm only,
\item {\tt double CMSENE } - Center of mass system energy,
                           used in reconstruction of $x_1, x_2$, 
                           the arguments for PDF distributions.
\end{itemize}

There are two random number generators used in typical applications. If re-decay 
of $\tau$ leptons by {\tt Tauola} is requested, then one may need to re-seed
its native random generator, especially if parallel generation is performed.
For this end, the sequence \\
\hskip 1 cm {\tt \small int ijklin=..., int ntotin=..., int ntot2n=...;}\\
\hskip 1 cm {\tt \small Tauola::setSeed(ijklin,ntotin,ntot2n);} \\
or
{\tt \small  Tauola::setSeed(time(NULL), 0, 0);} \\
should be considered. \\
The {\tt TauSpinner} algorithm for attribution of $\tau$ helicities
use no-sophisticate random number generator. It can be replaced with the command \\
\hskip 1 cm {\tt \small Tauola::setRandomGenerator( randomik );}

\noindent
Other initialization methods, specific only for some  applications 
will be presented in sub-appendices.
\vskip 2mm
  \centerline{\bf Event loop }
\vskip 2mm
In a loop, for each consecutive  event read from the datafile\footnote{
  In most cases event record format of  {\tt HepMC}  is
  used as an input.}, the four momenta and {\tt pdgID} 
of the  outgoing $\tau$ leptons and their\footnote{Lists named {\tt tau1\_daughters, tau2\_daughters} are needed.} decay products
are retrieved.
The intermediate state boson $X$ has to be known too (from the content of the event record or set manually).

For each  event, information can be read from datafile with the  help of \\
\hskip 1 cm {\tt \small int status = readParticlesFromTAUOLA\_HepMC(input\_file, X, tau, tau2, tau\_daughters, tau\_daughters2); }\\
method, present in file \\
 {\tt /TauSpinner/examples/read\_particles\_from\_TAUOLA.cxx} \\
but we assume that user will replace ti with  his own, matching the format of the datafile of the actual use. 

\vskip 2mm
  \centerline{\bf Single event analysis }
\vskip 2mm
On the basis of the intermediate boson $X$  {\tt pdgID} algorithm decides
which matrix element version is used
for calculating $R_{i,j}$ (and production weight as well). The four momentum of $X$ does not need to be equal to the sum
of the outgoing $\tau$ leptons (or $\tau$ lepton plus $\nu_\tau$ in case of $W^\pm$ or $H^\pm$ mediated processes).
The non-conservation is expected, and will be attributed to the presence of  final state emitted photons (prior to $\tau$ decays).    
Then, the weight defined as in Eq.~(\ref{eq:wtspin}) is calculated \\
\hskip 1 cm  $wt_{spin}$ {\tt \small  = calculateWeightFromParticlesH(X, tau, tau2, tau\_daughters,tau\_daughters2); }\\
If there is only a single $\tau$ among $X$ decay products then it is assumed that
{\tt tau2}=neutrino\\
\hskip 1 cm  $wt_{spin}$ {\tt \small  = calculateWeightFromParticlesWorHpn(X, tau, tau2, tau\_daughters); }\\
The formula Eq.~(\ref{eq:wtspin}) is used in both cases, for the second one
where instead of $\tau$, the $\nu_\tau$ is present, a  polarimetric 
vector  $(1,0,0,0)$ is used.


The methods prepared for $2 \to 4$ processes have similar input parameters,
only the four momenta corresponding to outgoing partons (jets) are added.
The event can be read from datafile, e.g. with the help of \\
 \hskip 1 cm {\tt \small int status = read\_particles\_for\_VBF(input\_file,p1,p2,X,p3,p4,tau1,tau2,tau1\_daughters,tau2\_daughters); }\\
and later the weight is calculated  \\
\hskip 1 cm  $wt_{spin}$ {\tt\small   = calculateWeightFromParticlesVBF(p3, p4, X, tau1, tau2, tau1\_daughters, tau2\_daughters); }
\vskip 2mm
  \centerline{\bf Auxiliary information  }
\vskip 2mm
As a by-product of preparing input for calculation of spin weight, Eq.~(\ref{eq:wtspin}), auxiliary
physics information is available and can be retrieved. This information is available through the following {\tt get} functions
that can be called after each {\tt calculateWeightFrom...} function call. \\
\hskip 1 cm  {\tt double wtnonSM = getWtNonSM()  } - returns
production weight $wt_{prod}$  given by. Eq.~(\ref{eq:prod}), \\
\hskip 1 cm  {\tt double WTp = getWtamplitP();  }  \hskip 1 cm  {\tt double WTm = getWtamplitM() } return
weights $wt_{decay}^{\tau^+}$,  $wt_{decay}^{\tau^-}$ of Eq.~(\ref{eq:dec}), corresponding to variation of $\tau^\pm$ decay matrix elements.
Finally \\
\hskip 1 cm  {\tt  double getTauSpin() } \\
returns   the $\tau$ lepton helicity, attributed stochastically from the  $p^z_\tau$  probability given by Eq.~(\ref{eq:probl}), which is calculated from
production and $\tau$ decay products kinematics. 
Note that such attribution is often subject of a 
further approximation than used for spin weight calculation:  transverse spin
effects are then  not included, and 
 100\% spin correlation is assumed. That is why, providing
 helicity only for the first $\tau$ is sufficient. \\
 The {\tt getEWwt(), getEWwt0()} methods are described in Appendix~\ref{app:electroweak}

 Before we explain further details of examples, let us list in the
 Table~\ref{Tab:methods}, all weights and their  methods,
and where they are adressed in Appendix.

\begin{table}

\vspace{2mm}
\begin{center}                               
\begin{tabular}{|c|c|c|c|}
\hline \hline
Weight           & Method      & Equation & Details \\
\hline 
$wt_{spin}$       & {\tt calculateWeightFromParticlesH}   &        &  \\
                 & {\tt calculateWeightFromParticlesVBF} &  (\ref{eq:wtspin})   & App.~\ref{app:jj} \\
                 & {\tt calculateWeightFromParticlesHpn} &        &  \\
\hline
                 & {\tt getTauSpin } &  (\ref{eq:probl})      & Sec.~\ref{sec:negl}  \\
\hline
$wt_{prod}$       & {\tt getWtNonSM()}        & (\ref{eq:prod})     & App.~\ref{app:ME} \\
$wt_{decay}$      & {\tt getWtAmplidP}, {\tt getWtAmplidM}  & (\ref{eq:dec})      & App.~\ref{app:rchl} \\
$wt_{EW}$        & {\tt getEWwt()}, {\tt getEWwt0()}  &       & App.~\ref{app:electroweak} \\
\hline 
\end{tabular}
\end{center}
\caption{ Weights of {\tt TauSpinner}, methods of calculation and Appendices
  explaining examples of their use.
 \label{Tab:methods} }
\end{table}

\subsection{ Longitudinal spin effects in $2 \to 2$ processes}\label{app:bench}
From the first version of {\tt TauSpinner} \cite{Czyczula:2012ny}, the purpose of calculating
spin weight $ wt_{spin}$ given by formula (\ref{eq:wtspin}),  was to evaluate 
how spin correlations between produced $\tau^+$ and $\tau^-$ affect
different observables.   
Applications for the cases of $Z, \gamma^*,mW^\pm, H, H^\pm$ mediated processes
were prepared.
Longitudinal spin effects were studied only. 
Other than $t,z$ components of the $R_{i,j}$ matrix, Eq.~(\ref{eq:wtspin}),
 were set to zero. Source of the paper \cite{Czyczula:2012ny} is  available in directory {\tt Tauola/TauSpinner/paper}.

Introduced at that time methods form the core of the {\tt TauSpinner} code.
Later, more refined matrix elements and accompanying  methods were 
introduced. The general principles remained unchanged.
The parameters of the Born amplitudes  in case of the Higgs mediated process can be modified with: \\
\hskip 1 cm  {\tt \small  void setHiggsParameters(int jak, double mass, double width, double normalization)}. \\
At the time, these parameters were not yet known experimentally.
For the Drell-Yan process the defaults of {\tt Tauola} \cite{Davidson:2010rw} are used.

Detailed
discussion of examples is presented with multitude  plots of
Ref.~\cite{Kaczmarska:2014eoa}. Option depicted by the initialization parameter 
{\tt Ipol} necessary to adopt to the level of spin effects embeded 
in, to be reweighted events, is discussed. Figures for this paper 
and the paper itself are stored in  {\tt TAUOLA/TauSpinner/examples/applications/}.


\subsection{ Beyond Standard Model Higgs and Drell-Yand $2 \to 2$ processes}\label{app:ME}
With Ref.~\cite{Banerjee:2012ez} re-weighting with the help of the production 
matrix element was introduced and $wt_{prod}$ of Eq.~(\ref{eq:prod}) became available.
The methods\\
\hskip 1 cm  {\tt setNonSMkey(1);}\\
\hskip 1 cm  {\tt double getWtNonSM();}
\\
were introduced. The facility was prepared for handling the `new' matrix
element  introduced by the user.
The methods \\
\hskip 1 cm  {\tt  set\_nonSM\_born( nonSM\_adopt );}
\hskip 1 cm  {\tt  set\_nonSM\_bornH( nonSM\_adoptH );}
\\
were provided. Numerical  values of the coupling constants for the default 
matrix elements of {\tt TauSpinner} can be  changed in that way as well.
We will exploit  that point in context of $2\to 4$ processes. 

The  user provided matrix elements can be defined for the
Higgs mediated processes or for the Dell-Yan ones. The algorithm will make 
a choice which one to use, on the basis of {\tt pdgID}  of the $X$.
With $wt_{prod}$ one can 
change  the initialization parameters of the Standard Model amplitudes as well.

If no user provided matrix elemens are attributed,
program will print warning and {\tt exit(-1)} command will be activated.

\subsection{Examples for studies of Higgs parity observables.}\label{app:CP}
Example program and reference numerical results
for CP sensitive observables as of  Ref. \cite{Przedzinski:2014pla} are 
 given in directory\\ {\tt Tauola/TauSpinner/examples/CP-tests}.  The $wt_{spin}$ is used.\\
Sub-directories with names starting with {\tt H-} and {\tt Z-},
are respectively for cases where the signal (Higgs boson) or background
(Drell-Yan, $Z$-production) 
distributions are  studied. The
second
 part of the directory name, denotes  decay channels of $\tau$ leptons,
{\tt pi} for $\tau \to \pi \nu$ and  {\tt rho} for $\tau \to \pi \pi^0 \nu$.

Sub-directory 
{\tt generate-datafiles} includes the program
 to read and generate $\tau$ decays 
for events stored in input {\tt Pythia HepMC}  files. 
%
The {\tt README} file with instructions is prepared, but was not updated now
in every detail.
Source of the paper \cite{Przedzinski:2014pla} is in directory: {\tt TAUOLA/TauSpinner/paper-CP}.

\subsubsection{CP Higgs}\label{app:CP-h}
From the algorithmic point of view
introduction of CP-sensititive transverse spin correlations
into $R_{i,j}$ matrix was straightforward and completed 
in Ref.~\cite{Przedzinski:2014pla}. The task was simple, because 
due to the zero spin of the Higgs there is no dependence on the
Higgs production process.
To activate transverse spin correlations, 
one should initialize one of the below:
\begin{itemize}
\item
{\tt setHiggsParametersTR(-1.0, 1.0, 0.0, 0.0);} for scalar Higgs, 
\item
{\tt  setHiggsParametersTR( 1.0,-1.0, 0.0, 0.0);} for pseudo-scalar Higgs,
\item
{\tt  double theta = 0.2;} for mixed parity state choose {\tt theta} for mixing angle and then \\
{\tt setHiggsParametersTR(-cos(2*theta),cos(2*theta),
-sin(2*theta),-sin(2*theta));} \\

\end{itemize}
For the code,
see {\tt TAUOLA/TauSpinner/examples/CP-tests } and its sub-directories
{\tt H-pi } and  {\tt H-rho } and for an example of results obtained with
the methods Refs.~\cite{Jozefowicz:2016kvz,Barberio:2017ngd}.

\subsubsection{Transverse spin correlations in the background; $2 \to 2$ Drell-Yan process}\label{app:CP-z}
With the help of the tables of electroweak corrections, the transverse spin effects of
$Z/\gamma^*$ mediated processes can be introduced into the sample. 
As default those terms are switched off. To activate
as explained  in Ref.~\cite{Przedzinski:2014pla},
initialization method \\
{\tt setZgamMultipliersTR(1., 1., 1., 1. );} has to be used.\\
User may be interested to introduce transverse spin correlations in part,
then some of these coefficients can be set to zero or even one can reverse their sign, by setting
multipliers to $-1.$
If any component of transverse density matrix for Drell-Yan has been turned on
using {\tt setZgamMultipliersTR}, spin weight $wt_{spin}$ will also
take into account transverse spin correlations $R_{i,j}, i,j=x,y$
(calculated from tables {\tt  table1-1.txt table2-2.txt}
which must be stored in the same directory as the program) by calling\footnote{
The {\tt S,cost} denote invariant mass squared of the
$\tau$-lepton pair as well as $\cos{\theta}$ (of the $2\to 2$ scattering angle)
and {\tt ID} the  identifier for the incoming parton flavour, all for the
parton level kinematic configuration prepared by {\tt TauSpinner}.
To get the actual values interpolations between entries of tables is used.
}
   sequence\\
{\tt Tauolapp::TauolaParticlePair pp;} \\
{\tt pp.recalculateRij(ID,tau\_pdgid,S,cost);}

\noindent
For the code,
see {\tt TAUOLA/TauSpinner/examples/CP-tests } and its sub-directories {\tt Z-pi } and  {\tt Z-rho}.

\subsection{Electroweak corrections} \label{app:electroweak}
In  Ref.~\cite{Przedzinski:2014pla} it is explained how  tables
featuring 
electroweak one loop effects can be installed into  weight 
calculations. At present, only electroweak corrections embedded as in the tables
\\ {\tt TauSpinner/examples/CP-tests/Z-pi/table1-1.txt},  \\
{\tt TauSpinner/examples/CP-tests/Z-pi/table2-2.txt} \\ are available.
These results do not include lineshape corrections, that is multiloop corrections
to vacuum polarization diagrams for quarks. That is why, they are of the restricted
practical use, except may be for the forward-backward asymmetry simulation,
unless replaced with the better ones, originating e.g. from the
{\tt DIZET} library of {\tt KKMC} program \cite{Jadach:1999vf} of the LEP time. 

If tables are present in the 
directory of 
the executable, then they will be read in and used.
Then, if \\
{\tt   double wt1=Tauola::getEWwt();}\\ {\tt double wt0=Tauola::getEWwt0();}\\
is executed, one can  also
 improve the distribution and normalization of the cross section
 with the {\tt wt1/wt0} factor added in the $wt_{prod}$  event weight
 and not only take transverse spin effects discussed above.

\subsection{Drell-Yan $\tau\tau$ pair production with two jets}\label{app:jj}
Necessary methods were introduced/discussed in Ref.~\cite{Kalinowski:2016qcd}:
it turned out of interest to extend calculation of matrix elements to the case where four-momenta of two  
jets accompany outgoing $\tau$ lepton pair, are also used explicitely.
This option is invoked, see {\tt TAUOLA/TauSpinner/examples/example-VBF}
by call to\\
\hskip 1 cm  {\tt\small  WT = calculateWeightFromParticlesVBF(p3, p4, X, tau1, tau2, tau1\_daughters, tau2\_daughters);} \\
instead of discussed previously \\
\hskip 1 cm  {\tt\small WT = calculateWeightFromParticlesH(...). }\\

The matrix elements replacement necessary for the calculation of
the corresponing $wt_{prod}$
is set by \\ {\tt TauSpinner::set\_vbfdistrModif(vbfdistrModif);}\\ 
and then initialized with \\
  {\tt  vbfinit\_(\&ref,\&variant); }
\\ 
For the non-standard calculation it is then
available with \\
{\tt  setNonSMkey(1);}.
\\
In~\cite{Kalinowski:2016qcd} it was found interesting to use
this interface to redefine initialization of the parameters
in the default matrix elements.
Distinct electroweak schemes can
be initialized for standard and non-standard calculations. 
Variants for  running $\alpha_S$ can be also distinct for standard and non-standard calculation.
Note that the default for non-standard calculation is to use the same set of
matrix elements as for Standard Model calculation. The electroweak
scheme can be nonetheless different (the one corresponding to {\tt variant}).
Also option for distinct scale in  calculation of PDFs is prepared.
 It can be set with \\
{\tt setPDFOpt(QCDdefault,QCDvariant);}
respecively for the default and non-standard calculation. Also,
new parametrization of running strong coupling can be introduced with this
interface.
The default function
can be replaced  
with the user own:\\
{\tt TauSpinner::set\_alphasModif(alphasModif);}.
These options were already explored in Ref.~\cite{Kalinowski:2016qcd}.

\subsection{Spin2}\label{app:spin2}

So far we have discussed production weight $wt_{prod}$ as a tool to reweight
to slightly modified initialization parameters of the Standard Model
production model. 
As demonstrated  in Ref.~\cite{Bahmani:2017wbm}, matrix elements calculated for
the non-standard type (eg. for non-standard resonance) may differ
substantially from the default-standard 
mode ones.  Invoking \\
{\tt TauSpinner::set\_vbfdistrModif(SPIN2::spin2distr);} will activate user function, as of our example, the 
{\tt SPIN2::spin2distr(...,KEY)}. In this case, user matrix element prototype invokes a whole library which  need initialization,
see {\tt TAUOLA/TauSpinner/examples/example-VBF/SPIN2}.
Its organization and initialization is quite similar to the code of the {\tt TauSpinner} library of
$\tau\tau j j$ processes, but it is nonetheless completely independent.
The  
{\tt spin2init\_(\&ref,\&variant);}
\\
has the same parameters as the {\tt vbfinit\_}, but not all of its 
functionality is at this moment active.  The parameter {\tt ref} is dummy
and all constants and coupling are independently initialized, in particular
hard coded own value for  
not-running $\alpha_s$ is used. The parameter {\tt variant} can be used to 
vary electroweak scheme though.

One should bear in mind, that re-weighting of matrix element require 
spin weight factor {\tt WTME} to be included with the $ wt_{prod}$, it has to be calculated as follows:
{\small 
\begin{verbatim}
setNonSMkey(0);  // to calculate spin weight of Standard Model
double WT0 = calculateWeightFromParticlesVBF(p3, p4, X, tau1, tau2, tau1_daughters, tau2_daughters);
setNonSMkey(1);  // e.g. for use of SPIN2 ME and calculate all weights
WT = calculateWeightFromParticlesVBF(p3, p4, X, tau1, tau2, tau1_daughters, tau2_daughters);
double WTME=getWtNonSM();        // matrix element weight
       WTME=WTME*WT/WT0;         // factor to take into account spin correlations of tau-tau pair decays
\end{verbatim}
}

\subsection{Changing matrix elements of $\tau$ decays.}\label{app:rchl}

Ambiguity of choice for the hadronic currents used in simmulations
may be a source of systematic 
errors, which is of the interest to be estimated. An attempt for such
estimation is shown in an example  \\
{\tt TAUOLA/TauSpinner/examples/CLEO-to-RCHL/tau-reweight-CLEO-to-RCHL.cxx} 

The calculation  with two versions of hadronic currents is possible
without any intervention into the code of  libraries and interchanged 
thanks to the method  {\tt Tauola::setNewCurrents( int key);} where
{\tt key=0} or {\tt key=1} is possible. Once hadronic current model is set,
the usual weight calculation
can be performed 
and
{\tt double getWtamplitP(); double getWtamplitM();}  can be used to obtain matrix
elements squared respectively for
$\tau^+$ and $\tau^-$.   Ratio of the results for the  two versions of
hadronic currents will constitute
the  $wt_{decay}^{\tau^+}$ and  $wt_{decay}^{\tau^-}$ weights.

\subsection{Not published tests programs}\label{app:other}
Set of selected test programs, coded without {\tt MCTester} user analysis helper,  
are collected in  directory \\ {\tt TAUOLA/TauSpinner/examples/testsERW}.

\subsection{Function sigborn}\label{app:sigborn}

Input parameters are {\tt pdgID} of incoming partons, virtuality squared of the resonance $Q^2$ 
and $\cos \theta$ of the outgoing lepton, the intermediate  particle identifier is not passed as an argument.
Appropriate choice of the function used is  performed by the {\tt TauSpinner} method
calculating spin and production weight.

In case of the SM Higgs boson production, assumed is $gg \to H$ production process only, the formula for the 
lowest-order cross-section summed over all $\tau$ helicity configurations using wrapper to {\tt disth} function from 
{\tt Tauola } code read as follows
\begin{eqnarray}
\sigma_{Born} &=& \frac{1}{\pi} \frac{m_{h}^3 \cdot \Gamma_h}{(Q^2 - m_h^2)^2 +  m_h^2 \cdot \Gamma_h^2}  \nonumber \\ 
             &&  (  disth(Q^2, \cos \theta, 1, 1)  + disth(Q^2, \cos \theta, 1, -1) \label{eq:sigb}\\
             &&  +  disth(Q^2, \cos \theta,-1, 1) +  disth(Q^2, \cos \theta, -1, -1)).\nonumber  
\end{eqnarray} 

In case of BSM Higgs boson following formula is coded
\begin{eqnarray}
\sigma_{Born} &=& \frac{1}{\pi} X_{norm} \frac{m_{h} \cdot \Gamma_h}{(Q^2 - m_h^2)^2 +  m_h^2 \cdot \Gamma_h^2}  \nonumber \\ 
             &&   (  disth(Q^2, \cos \theta, 1, 1)  + disth(Q^2, \cos \theta, 1, -1) \\
             &&  +  disth(Q^2, \cos \theta,-1, 1) +  disth(Q^2, \cos \theta, -1, -1)) \nonumber 
\end{eqnarray} 
where $ X_{norm}$ (in units of GeV$^{2}$) stands for  normalisation factor, see Ref.~\cite{Banerjee:2012ez}. 

In case of Drell-Yan process, respective formula reads, using wrapper to {\tt t\_born} function from {\tt Tauola} code.
\begin{eqnarray}
\sigma_{Born} &=&  \frac{1}{ Q^2 \cdot 123231}  \nonumber \\
             &&   (  t\_born(0, Q^2, \cos \theta, 1, 1)  + t\_born(0, Q^2, \cos \theta,  1, -1) \\
             &&   +  t\_born(0, Q^2, \cos \theta,-1, 1) +  t\_born(0, Q^2, \cos \theta, -1, -1)) \nonumber 
\end{eqnarray}
where $Q^2$ is the virtuality of the resonance, $\cos \theta$ is the cosine  of angle between $\tau^+$ and the first beam
and last two arguments are helicity states of $\tau^+$ and $\tau^-$ respectively. 
The 123231 = $ \sim 2 \cdot \pi  / \alpha^2_{QED}$ represent
normalisation factor.
Details of normalization were not studied so far, this was out of scope for
the studied applications, as in the $wt_{spin}$  ratios of  $\sigma_{Born}$ were
used and for  $wt_{prod}$ required precision was sufficent, in cases when
required precision was higher the ratio
of  $\sigma_{Born}$ was used for  $wt_{prod}$ as well.

\begin{table}
\vspace{2mm}
\begin{center}                               
\begin{tabular}{|l|r|r|}
\hline \hline
Parameters           & {\tt initwk }     & {\tt DISTJKWK} \\
\hline 
$\sin\theta_W^2$     & 0.23147      &  0.2315 \\
$m_e$                & 0.511 MeV   &  -- \\
$m_Z$                & 91.1882 GeV &  91.187 GeV\\
$\Gamma_Z$           & 2.4952 GeV  &  2 GeV \\
$m_{\tau}$            & 1.77703 GeV &    --     \\
$\alpha_{QED}(m_{Z})$  &             & 1./128.  \\
$G_F$                &             & $1.1667 \cdot 10^{-5}$  \\
\hline 
\end{tabular}
\end{center}                               
\caption{ Setting of EW parameters in {\tt initwk} routine for effective Born of {\tt Tauola} and in routine 
  {\tt DISTJKWK} of effective Born used in Ref.~\cite{Banerjee:2012ez}.
  Note that the parameters do not match. The purpose of electroweak
  parameters in  {\tt DISTJKWK} was to relate New Physics amplitudes
  with the Standard Model ones, its own calculation scheme of quite rough
  Standard Model initialization was used.
 \label{Tab:initEW} }
\end{table}

The electroweak  parameters are initialised with the setting as in Table~\ref{Tab:initEW}, 
using wrappers to  {\tt initwk}\_ function from {\tt Tauola} library code. 
In Appendix~\ref{app:electroweak}, it is explained how to activate
electroweak one loop corrections, of physics content as  described
in Refs.~\cite{Andonov:2008ga,Andonov:2004hi} into effective Born angular distributions.
Appendices C.7 and  D of Ref.~\cite{Davidson:2010rw} document how
the scheme of these electroweak
effects calculation can be changed.
The corresponding example code is provided in
{\tt TAUOLA/TauSpinner/examples/testsERW/ex4}


\end{document}